%% file: ttbarmssm.tex
\documentclass[en]{paper}

\usepackage{axodraw}
\usepackage{fmslash}

\newcommand{\Fermion}{F}
\newcommand{\Vector}{V}
\newcommand{\Scalar}{S}
\newcommand{\Quark}{q}

\newcommand{\cconj}{\mathsf{c}}

\begin{document}
\bibliographystyle{utphys}

\input{title.tex}
\input{intro.tex}
\input{hel_mat_elem.tex}
\input{prototypes.tex}
\input{results.tex}

\input{summary.tex}

\clearpage
\bibliography{ttbarmssm}
\end{document}

%% file: title.tex
\begin{flushright}
{SHEP-07-19}\\ 
{\today}
\end{flushright}

\vspace*{2.0truecm}
\begin{center}
{\Large \bf MSSM Effects in Top-antitop Production at the LHC.}
\\[1.5cm]
{\large D.A. Ross,  and M. Wiebusch}\\[0.15 cm]
{\it School of Physics and Astronomy, University of Southampton}\\
{\it Highfield, Southampton SO17 1BJ, UK}\\[0.25cm]
\end{center}
\vspace*{0.5cm}
\begin{abstract}
\noindent
We report on a calculation of the effects of supersymmetry on
the cross-section for $t-\bar{t}$ production at LHC. A numerical study is
carried out for the ten benchmarks of the Snowmass accord. It is found that
the higher order effects involving supersymmetric particles in internal
 loops can be as high as 6\%, both for the cross-section and the 
(parity even) helicity asymmetry, for one particular benchmark.
For other benchmarks smaller but nonetheless observable corrections are found.

\end{abstract}
\vspace*{0.4cm}
\centerline{Keywords:~{Hadron Colliders, Supersymmetry, 
Higher-order calculations.}}

%% file: intro.tex
\newpage

\section{Introduction}
\label{sec:intro}

For nearly 35 years, supersymmetry (SUSY) has been an attractive theory
 in particle physics. At the most theoretical level, it permits 
the construction of string theories which do not
contain tachyonic states, and at the phenomenological level it offers
an explanation for the naturalness of the hierarchy through its
reduced ultraviolet divergences, as well as providing
resolutions of several puzzles arising in standard models of
cosmology. It also gives rise to a correction to the running of the couplings,
so that the strong, weak, and electromagnetic interactions can unify
at some Grand Unified (GUT) scale.

However, to date there has been no reliable evidence that this theory
describes Nature, so that if SUSY is indeed realised in nature, it must
 be broken at a scale higher than that reached in accelerator 
experiments conducted up to now. If the theory is to be effective
in providing a solution to the naturalness of the hierarchy problem, then
 the supersymmetry breaking scale cannot be much more than about 1 TeV.
This is also the scale of SUSY breaking which leads to unification
of couplings. This implies 
 that, with the exception of some hidden corners of parameter space,
SUSY is expected to be discovered at the forthcoming LHC. 

Clearly the most dramatic manifestation of SUSY would be the production
and identification of supersymmetric partner particles such as 
the spin-$\frac{1}{2}$ charginos or neutralinos, or evidence that at
sufficiently high energies hadrons display behaviour consistent with
the existence of squarks or gluinos in the sea. Nevertheless, the existence
of supersymmetry will have  indirect but measurable
effects on the cross-sections for the production of Standard
Model (SM) particles. The LHC is expected to achieve sufficient
integrated luminosity such that it will be possible to determine these
cross-sections with sufficient accuracy to be able to detect the effects
of higher order corrections coming from loops of supersymmetric
particles. The loops can give rise to a significant correction
to the production cross-sections even below the threshold for the
production of the supersymmetry particles themselves, so that
hints that some new physics is imminent can be deduced before the
threshold energies are actually reached. Above these thresholds, 
differential cross-sections
with respect to suitably chosen variables can display structures which
can be used to determine masses of some of the 
scalar particles of the MSSM.

In this paper, we consider the influence of the minimal supersymmetric 
extension of the Standard Model (MSSM) to the production
of top-antitop $t-\bar{t}$. At LHC, after an integrated luminosity of
100 $\mathrm{fb^{-1}}$, one expects around $2 \times 10^7$ such events,
 so that small corrections should be easily identifiable, not only at the
level of the total production cross-section, but also for
differential cross-sections with respect to the transverse momentum,
$p_T$ of the $t$-quark, and also with respect to the invariant
mass $M_{t\bar{t}}$, of the  $t-\bar{t}$ pair.
Both of these are expected to be measured accurately at LHC.
The latter variable is one for which differential cross-sections
display a discernible structure as one crosses various thresholds.
Apart from statistical errors, there will be larger systematic errors
arising form uncertainties in the incident beam flux and in the
parton distribution functions (PDF's). Such systematic errors are
 cancelled in the ratio of helicity asymmetries to total cross-section,
defined in Eqs.(\ref{eq:res:asymmetries}) and (\ref{eq:res:ratios}).
Although the helicities of individual $t$- or $\bar{t}$-quarks cannot
be determined on an event-by-event basis, their distributions can be
 inferred from the angular distributions of the decay products
(see \cite{ttpol}). The corrections to these asymmetries can be as high
as 5 \%.

Although the MSSM has very few extra parameters in the supersymmetry conserving
sector, the supersymmeric breaking terms can introduce 105 parameters in
addition to the 19 parameters of the SM.  Clearly, it is impossible to analyse
the complete space of these parameters.  Nevertheless, we have therefore
organised our calculation so that software is available to calculate
differential cross-sections with all possible helicity configurations and any
given set of SUSY parameters.  We have done this by working in terms of helicity
matrix-elements and setting up ``prototype graphs'' which can then be included
in the calculation with any combination of couplings and internal masses.  This
provides maximum flexibility for the extraction of total or differential
cross-sections, helicity asymmetries, etc. as well adaptability to other SUSY
models. The library was also designed to scan efficiently over a large number of
parameter sets. This was achieved by keeping, for one set of kinematic
parameters, a table of all required Veltman-Passarino functions in memory, thus
avoiding to re-calculate these functions for each new set of model parameters.
This can reduce the computation time by up to 60\% and makes scans over one
or even two-dimensional parameter spaces feasible.

The number of independent SUSY parameters is greatly reduced if one considers
supersymmetry scenarios which are consistent with super-gravity or in which SUSY
breaking is either gauge-mediated or mediated by the Weyl anomaly. At the
Snowmass meeting of 2002, \cite{Snowmass02}, a collection of ten typical models
was considered and sets of SUSY parameters for these models were generated. For
convenience, we have investigated the effects of SUSY corrections to $t-\bar{t}$
production for these ten parameter sets.  We find considerable variation in the
magnitudes of the corrections from these different parameter sets. Conversely,
this means that accurate measurement of the $t-\bar{t}$ production cross-section
can be used as a tool to help identify the correct set of SUSY parameters.

At sufficiently high (partonic) energies, the SUSY corrections
to  $t-\bar{t}$ production are expected to be dominated by
single and double  logarithms of incoming parton energy divided by
the SUSY breaking scale, $M_{SUSY}$. The determination
of these logarithms is independent of the SUSY parameter set,
with the exception of  $M_{SUSY}$ and the ratio, $\tan\beta$,
of the vacuum expectation values of the two Higgs doublets,
and the calculation is simplified by the fact that the mixing of
various supersymmetric particles to form mass eigenstates has no effect
on these logarithms. The logarithmic contributions have been
calculated  by Beccaria et. al. \cite{Beccaria04}.
One may have expected that it would have been possible to express the
 entire SUSY correction in terms of these logarithms plus a constant off-set,
which depended on the SUSY parameter set. We have compared our results
 with those of ref.\cite{Beccaria04} and although it is indeed the case that
our results agree with these logarithms plus a constant off-set at sufficiently
high partonic energies, this approximation is found to be unsuitable at
typical partonic energies which will be reached at LHC, and the entire
calculation is required for a reliable 
prediction of the cross-sections at LHC.

The structure of this paper is as follows: In {\it section} \ref{sec:hel_mat_elem}
we discuss the general method for the extraction of the
 above-mentioned helicity matrix-elements at the partonic level
from a general Feynman graph.
In {\it section} \ref{sec:pro} we
 list all the prototype graphs and indicate which
supersymmetric particles can contribute for each of the prototypes.
In {\it section}   \ref{sec:results} we discuss the results after
folding the partonic cross-sections with PDF's and show the results for
the ten Snowmass benchmark points. {\it Section} \ref{sec:summary}
presents some conclusions.

%% file: hel_mat_elem.tex
\section{Helicity Matrix Elements}
\label{sec:hel_mat_elem}
Because we wish to be able to discuss the total and differential
 cross-sections for given helicities of the $t$- and $\bar{t}$-
quarks as well as the asymmetries, we find it convenient to work
at the parton level for a generic process 
$$ p_a(\lambda_a) \, + \, p_b(\lambda_b) \ \to \ 
 p_1(\lambda_1) \, + \, p_2(\lambda_2) $$
 in terms of helicity matrix-elements
${\cal A}_{\lambda_a,\lambda_b,\lambda_1,\lambda_2}(E,\theta)$,
as functions of the partonic centre-of-mass energy, $E$ and scattering
angle $\theta$ (also in the centre-of-mass).

In the case of quark-antiquark annihilation, for
which we take the incoming quarks and antiquarks to be massless,
the helicities of the incoming 
partons are anti-correlated ($\lambda_a = -\lambda_b$),
and although it is possible for this initial helicity anti-correlation to be
violated at the one-loop level, the interferences at ${\cal O}(\alpha_s^3)$
or  ${\cal O}(\alpha_s^2\alpha_W)$ 
always respect this anti-correlation. On the other
hand, for the gluon fusion process there need be no
correlation between gluon helicities.

Once these helicity matrices have been determined, differential cross-sections
and asymmetries can be computed by convolution with the corresponding
parton distribution functions (PDF), summed over helicities or not, 
as appropriate.

A further potential advantage of the helicity matrix-element formalism,
although not currently applicable at LHC, is the determination of 
initial beam polarisation asymmetries, should it become possible
in the future to polarise these beams. For interactions of a parity violating
nature such as SUSY corrections to weak interactions, such asymmetries would be
immensely useful in identifying the parameters of the supersymmtery model.

The helicity amplitudes are obtained in two stages. In the first stage,
a set of coefficient functions, 
$a^{\{\ \alpha\}}_{\lambda_a,\lambda_b}(E,\theta)$, of complete set of
Dirac $\gamma$-matrices, is determined:
\begin{equation}
{\cal A}_{\lambda_a,\lambda_b,\lambda_1,\lambda_2}(E,\theta)
 \ = \ \sum_{\{\alpha\}}a^{\{\alpha\}}_{\lambda_a,\lambda_b}(E,\theta)
\bar{u}(p_1,\lambda_1) \Gamma^{\{\alpha\}} v(p_2,-\lambda_2) \label{hme1}
\end{equation}
where $\Gamma^{\{\alpha\}}$ are the matrices
\begin{eqnarray}
\Gamma_V^\mu &=& \gamma^\mu \nonumber \\
\Gamma_A^\mu &=& \gamma^\mu \gamma^5 \nonumber \\
\Gamma_T^{\mu\nu} &=& \sigma^{\mu\nu} \end{eqnarray} 
with associated projection operators $P^{\{\alpha\}}$
\begin{eqnarray}
P_V^\mu &=& \frac{1}{4} \left( \gamma^\mu + \frac{p_1^\mu}{m_t}\right)
 \nonumber \\
P_A^\mu &=&  \frac{1}{4} \left(
\gamma^5\gamma^\mu +  \gamma^5 \frac{p_1^\mu}{m_t}\right)\nonumber \\
P_T^{\mu\nu} &=& \frac{1}{8}\sigma^{\mu\nu} \end{eqnarray},
such that
$$ \mathrm{Tr} \left( \Gamma^{\{\alpha\}} P_{\{\beta\}} \right) \ = \ 
\delta^{\{\alpha\}}_{\{\beta\}}.$$
Note that the coefficient functions are independent of the helicities of
the $t$- and $\bar{t}$ quarks

For the basis vectors $e_0 \cdots e_3$, where
\begin{eqnarray}
e_0^\mu &=& \frac{1}{2E} \left( p_1^\mu+p_2^\mu \right) \nonumber \\
e_1^\mu &=& \frac{1}{2Ep\sin\theta}
 \left(-(p+E\cos\theta) p_1^\mu + (p-E\cos\theta) p_2^\mu
 + 2p\,  p_a^\mu \right)\nonumber \\
e_2^\mu &=& \frac{1}{2E^2p\sin\theta}
 \epsilon^\mu_{\nu\rho\sigma} p_1^\nu p_2^\rho p_a^\sigma \nonumber \\
e_3^\mu &=& \frac{1}{2p} \left( p_1^\mu-p_2^\mu \right)
 \nonumber \end{eqnarray}
($p$ being the magnitude of the three-momentum of the $t$-quark 
in the centre-of-mass frame),
the helicity matrix elements are given by
\footnote{These matrix-elements are defined up to an overall
phase, which may depend on the initial-state and final-state helicities.}
\begin{eqnarray}
{\cal A} &=& 2\left[ E \left(- a_V^1 +i \lambda_1 a_V^2 \right)
+i p \left(a_A^2 + i \lambda_1 a_A^1 \right) - i m_t \left(
a_T^{01}-i \lambda_1 a_T^{02} \right) \right] \delta_{\lambda_1,-\lambda_2}
\nonumber \\
 &+& 2\left[m_t \, a_A^0-p \, a_T^{12}-i \, E \,\lambda_1a_T^{03}-m_t
\, \lambda_1 a_V^3
\right]  \delta_{\lambda_1,\lambda_2} \label{hme} \end{eqnarray}

\begin{figure}[h]
  \centering
  \includegraphics{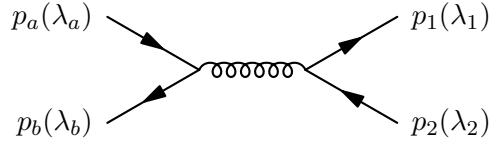}
  \caption{\label{hf1} Tree-level graph for $t-\bar t$ production from
    quark-antiquark annihilation.}
\end{figure}

Thus, for example, the non-zero coefficient functions for the
quark-antiquark annihilation process at the tree-level 
({\mrefs Figure \ref{hf1}}) are:
\begin{eqnarray}
a_V^1 &=& g_s^2 \frac{\cos\theta}{2E} 
\delta_{\lambda_a,-\lambda_b} 
\left( {\mbox{\boldmath $\tau$}} \otimes  {\mbox{\boldmath$\tau$}} \right)
\nonumber \\
a_V^2 &=& g_s^2 \frac{i\lambda_a}{2E} 
\delta_{\lambda_a,-\lambda_b}
\left( {\mbox{\boldmath $\tau$}} \otimes  {\mbox{\boldmath $\tau$}} \right)
\nonumber \\
a_V^1 &=& g_s^2 \frac{\sin\theta}{2E} 
\delta_{\lambda_a,-\lambda_b} 
\left( {\mbox{\boldmath $\tau$}} \otimes  {\mbox{\boldmath $\tau$}} \right),
\end{eqnarray}
where 
$\left( {\mbox{\boldmath $\tau$}} \otimes  {\mbox{\boldmath$\tau$}} \right)$
indicates the colour factor for a single gluon exchange. Inserting these
expressions for the coefficents into eq.(\ref{hme})
generates the helicity matrix-element
\begin{eqnarray}
{\cal A}_{\lambda_a,\lambda_b,\lambda_1,\lambda_2}(E,\theta) &=&
-g_s^2 \left( \lambda_a+\cos\theta\right)
\delta_{\lambda_a,-\lambda_b} \delta_{\lambda_1,-\lambda_2}
\left( {\mbox{\boldmath $\tau$}} \otimes  {\mbox{\boldmath $\tau$}} \right)
 \nonumber \\ & & 
 -g_s^2 \frac{m_t}{E} \lambda_1 \sin\theta \,
\delta_{\lambda_a,-\lambda_b} \delta_{\lambda_1,\lambda_2}
\left( {\mbox{\boldmath $\tau$}} \otimes  {\mbox{\boldmath $\tau$}} \right),
\end{eqnarray}

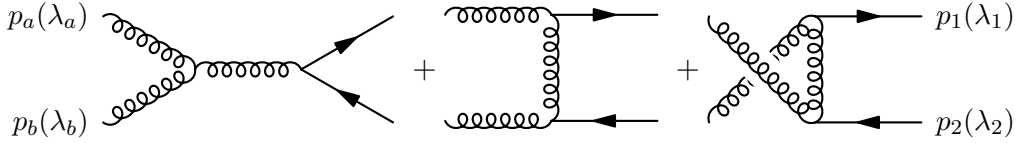
\begin{figure}[h]
  \begin{equation*}
     \mathalign{\input{hel-tree_s.tex}}
    +\mathalign{\input{hel-tree_t.tex}}
    +\mathalign{\input{hel-tree_u.tex}}
  \end{equation*}
  \caption{\label{hf2} Tree-level graphs for $t-\bar t$ production from gluon
    fusion.}
\end{figure}

whereas for the gluon fusion process ({\mrefs Figure \ref{hf2}}),
the non-zero coefficients are given by
\begin{eqnarray}
a_V^1 & = & \frac{g_s^2}{2E(E-p\cos\theta)} \tau^b \tau^a
  \left\{E \sin\theta \left(\cos\theta-2\right)
\delta_{\lambda_a,\lambda_b}+2p\sin\theta \cos\theta \right\} \nonumber \\
 & &  \ \ \ + \left[ \tau^a \leftrightarrow \tau^b, \ 
 \lambda_a \leftrightarrow \lambda_b, \ \theta \to (\pi+\theta) \right]
\nonumber \\
a_V^2 &=& \frac{ig_s^2}{2E(E-p\cos\theta)} \tau^b \tau^a
   p \, \sin\theta \, \lambda_a \delta_{\lambda_a,-\lambda_b}
  \ \ \ + \left[ \tau^a \leftrightarrow \tau^b, \ 
 \lambda_a \leftrightarrow \lambda_b, \ \theta \to (\pi+\theta) \right]
\nonumber \\
a_V^3 &=& \frac{g_s^2}{2E(E-p\cos\theta)} \tau^b \tau^a
  \left\{-E \cos\theta \left(\cos\theta-2\right)
\delta_{\lambda_a,\lambda_b}+2p\sin^2\theta  \right\} \nonumber \\
 & &  \ \ \ + \left[ \tau^a \leftrightarrow \tau^b, \ 
 \lambda_a \leftrightarrow \lambda_b, \ \theta \to (\pi+\theta) \right]
\nonumber \\
a_A^0 &=& \frac{g_s^2}{2E(E-p\cos\theta)} \tau^b \tau^a
  \lambda_a \left(p\cos\theta-E\right) \delta_{\lambda_a,\lambda_b}
 \nonumber \\ 
 & &  \ \ \ + \left[ \tau^a \leftrightarrow \tau^b, \ 
 \lambda_a \leftrightarrow \lambda_b, \ \theta \to (\pi+\theta) \right]
\nonumber \\
a_T^{12} &=&  \frac{g_s^2}{2E(E-p\cos\theta)} \tau^b \tau^a m_t \cos\theta
\delta_{\lambda_a,\lambda_b}
  \ \ \ + \left[ \tau^a \leftrightarrow \tau^b, \ 
 \lambda_a \leftrightarrow \lambda_b, \ \theta \to (\pi+\theta) \right]
\end{eqnarray}
leading to a tree-level helicity matrix-element
\begin{eqnarray}
{\cal A}_{\lambda_a,\lambda_b,\lambda_1,\lambda_2}(E,\theta)  & = & 
 \frac{g_s^2}{2E(E-p\cos\theta)} \tau^b \tau^a \nonumber \\ & \times & \Bigg\{
 m_t\left[2\lambda_1 \left((E+p)\cos^2\theta-2p-4E\cos\theta\right)
-2\lambda_aE
\right] \delta_{\lambda_1,\lambda_2} \delta_{\lambda_a,\lambda_b}
 \nonumber \\ & &  
\left[-2 E(E+p) \cos\theta\sin\theta+4E^2\sin\theta \right]
 \delta_{\lambda_1,-\lambda_2} \delta_{\lambda_a,\lambda_b}
 \nonumber \\ & & 
- 2 \lambda_1 p \, m_t \sin^2\theta 
  \delta_{\lambda_1,\lambda_2} \delta_{\lambda_a,-\lambda_b}
\nonumber \\ & &
-Ep\sin\theta \left(2\cos\theta+\lambda_1\lambda_a \right)
 \delta_{\lambda_1,-\lambda_2} \delta_{\lambda_a,-\lambda_b} \Bigg\}
\nonumber \\
& &   \ \ \ + \left[ \tau^a \leftrightarrow \tau^b, \ 
 \lambda_a \leftrightarrow \lambda_b, \ \theta \to (\pi+\theta) \right]
\end{eqnarray}

In the case of gluon fusion the contribution from any 
graph may be written in the form
$$ \bar{u}(p_1,\lambda_1) \Gamma v(p_2,-\lambda_2), $$
where $\Gamma$ is a sum of strings
 of $\gamma$-matrices with coefficients that
are proportional to couplings, internal and external fermion
masses and the Veltman-Passarino (VP) \cite{VP}
functions arising from the loop integrals. These VP functions have
 arguments that depend on the internal and external masses as well as
on the Mandelstam variables $s,t,u$. The coefficients
$a^{\{\alpha\}}$ are simply projected by
\begin{equation}
 a^{\{\alpha\}} \ = \ 
  \mathrm{Tr}\left( P^{\{\alpha\}} \Gamma \right) . 
\end{equation}

For the quark-antiquark annihilation process
there are two types of contributing  graphs:

The first type are graphs for which the fermion lines can be factorised
 into an initial quark line, $\Gamma^i$ and a final $t$-quark line, $\Gamma_f$.
Again these are sums of string of $\gamma$-matrices with coefficients that
are proportional to VP functions. In this case the coefficients
$a^{\{\alpha\}}$ are projected by
\begin{equation}
a^{\{\alpha\}} \ = \ \delta_{\lambda_a,-\lambda_b} \mathrm{Tr}
 \left( \Gamma_i \gamma \cdot v
 \frac{\left(1-\lambda_a \gamma^5\right)}{2} \right)
 \mathrm{Tr} \left(  P^{\{\alpha\}}\Gamma_f \right), \end{equation}
where $v^\mu$ is a  vector in the plane normal
to the incoming momenta $p_a$ and $p_b$ given by:
$$ v^\mu \ = \ 
\frac{1}{\sqrt{2} E \, p \, \sin\theta} 
 \left\{p_b \cdot p_1 p_a^\mu +p_a \cdot p_1 p_b^\mu -  p_a \cdot p_b p_1^\mu
 + i \lambda_a \epsilon^\mu_{\nu\rho\sigma} p_a^\nu p_b^\rho p_1^\sigma
\right\} $$

\begin{figure}[h]
  \centering
  \includegraphics{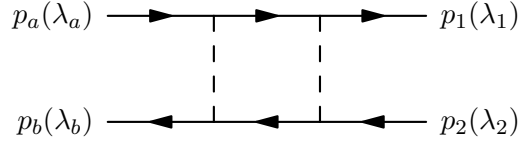}
  \caption{Gluino, neutralino, or chargino exchange
  contribution to  $t-\bar{t}$
  production from quark-antiquark annihilation}
  \label{hf5}
\end{figure}

The other type of graph is one in which the fermion lines do {\it not}
factorise into an initial fermion line and a final-fermion line,
but rather into an upper line, $\Gamma_u$, and a lower line, $\Gamma_d$,
in which the incoming quark and outgoing $t$-quarks are connected
by the exchange gluino, neutralinos, and charginos,is the $s$-channel.
It is graphs of this type that give rise to non-zero amplitudes
in the case where the incoming helicities are equal, but such
amplitudes do not interfere with the tree-level amplitudes. 
An example of such a graph is shown in {\mrefs Figure \ref{hf5}}.

For such graphs the coefficients
$a^{\{\alpha\}}$ are projected by
\begin{equation}
a^{\{\alpha\}} \ = \ \mathrm{Tr} \left( \Gamma_d \gamma \cdot v
\frac{\left(1-\lambda_a \gamma^5\right)}{2} \right) \Gamma_d
 P^{\{\alpha\}} \end{equation}

As in the case of most of the contributing graphs,
the expressions obtained from {\mrefs Figure \ref{hf5}} for the corrections to the
coefficients are too long and unwieldy to be reproduced here

\begin{figure}[h]
  \centering
  \includegraphics{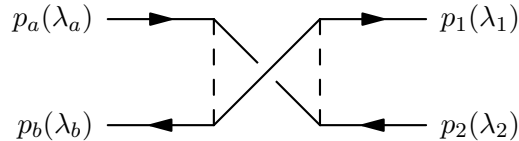}
  \caption{Graph involving the exchange of intermediate
    Majorana fermions}
  \label{hf6}
\end{figure}

Finally we note here that in several cases, the internal fermions exchanged
in the $s$-channel in such graphs may be neutralinos or gluinos,
which are Majorana fermions. In such cases supplementary graphs of the type
shown in {\mrefs Figure \ref{hf6}}
 need to be considered. Great care needs to be taken
in handling such graphs. The standard expressions for the propagators
of Majorana fermions in which a fermion propagates into a fermion
or an ant-fermion propagates into and anti-fermion, are ambiguous up to
a sign until the exact ordering of the fermions is determined. In order 
to ensure that this is effected in a consistent manner it is necessary
to determine the fermion ordering of the  term in
the Wick contraction corresponding to the graph under consideration.

A library has been constructed both in FORTRAN and C++ in which each
of the prototype graphs shown in the next section can be determined
numerically, as a function of the 
incoming energy, scattering angle, helicities,
couplings and internal masses. 
We have checked all prototype graphs by selecting different routings
 of the internal loop momenta and copmparing the results numerically.
The numerical values of
of the relevant VP functions are determined either using
the FF library \cite{ff} or LOOPTOOLs \cite{looptools}
These libraries can be found at \\
http://hep.phys.soton.ac.uk//hepwww/staff/D.Ross/susyttbar/

%% file: hel-tree_s.tex
\raisebox{-26.3045bp}{\includegraphics{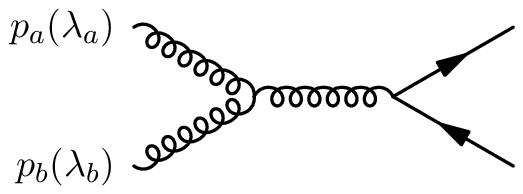}}

%% file: hel-tree_t.tex
\raisebox{-26.6667bp}{\includegraphics{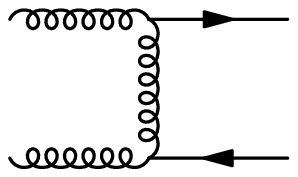}}

%% file: hel-tree_u.tex
\raisebox{-26.3045bp}{\includegraphics{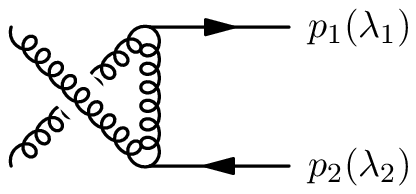}}

%% file: prototypes.tex
%
%
%
\section{Diagrams}\label{sec:pro}
%
%
In this section we list the diagrams needed for the computation of SUSY
contributions to polarised top-antitop production cross sections. To save space
we only draw the different \emph{topologies}. For further reference, each
topology has a label set in typewriter font. An asterisk behind the label
indicates that the crossed version of this diagram has to be included as well.
A double asterisk indicates that \emph{only} the crossed version is
needed. First or second generation quarks are labelled $\Quark$ and top quarks
are labelled $t$. Gluons are denoted as $g$ and always represented by curly
lines. Momenta and helicities are given in parentheses behind the label. For
example, $t(p,\lambda)$ denotes a top quark with (four-)momentum $p$ and
helicity $\lambda$. The generic scalar, vector and fermionic particles in each
topology are given uppercase labels $S$, $V$, $F$, etc. For each topology we
provide a list of MSSM particles that have to be substituted for the generic
ones. For the various MSSM particles we use the notations from \cite{Rosiek95}. However,
unless stated otherwise the generation indices $I$, $J$, etc.\ only run over the
first two generations. The third generation quarks are written explicitly as
$t\ (=u^3)$ and $b\ (=d^3)$.
%
%
\subsection{Tree-Level Diagrams}
%
\newcommand{\includediagram}[1]{%
  \parbox[t]{0.5\textwidth}{%
    \centering%
    \includegraphics{pro-#1.eps}
    \par
    \texttt{#1}%
  }%
}
\newcommand{\includediagrams}[2]{%
  \parbox[t]{0.5\textwidth}{%
    \centering%
    \includegraphics{pro-#1.eps}
    \par
    #2%
  }%
}
The following prototype diagrams contribute to $t\bar t$ production at
tree-level:
\par\bigskip
\noindent
\includediagrams{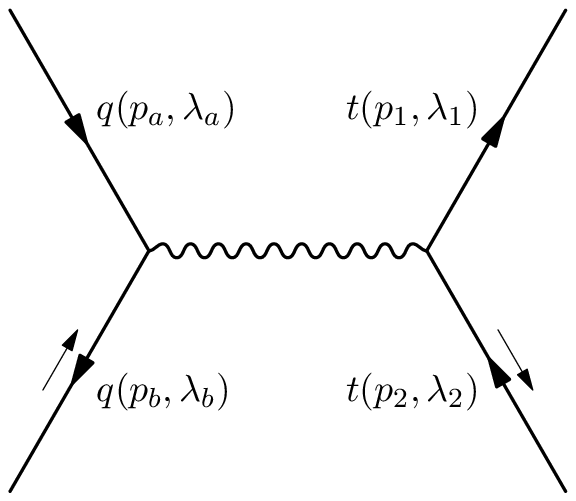}{%
  \texttt{Dqqbar\_sV\_tree}\\
  $\Vector = \gamma,Z,g$
}%
\includediagrams{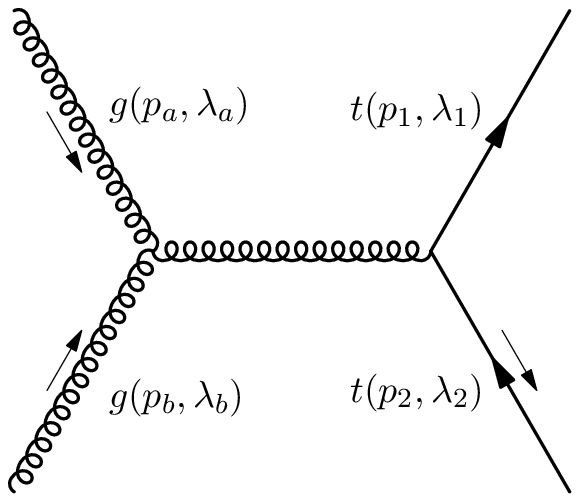}{\texttt{Dgg\_sG\_tree}}
\par\bigskip
\noindent
{\centering\includediagrams{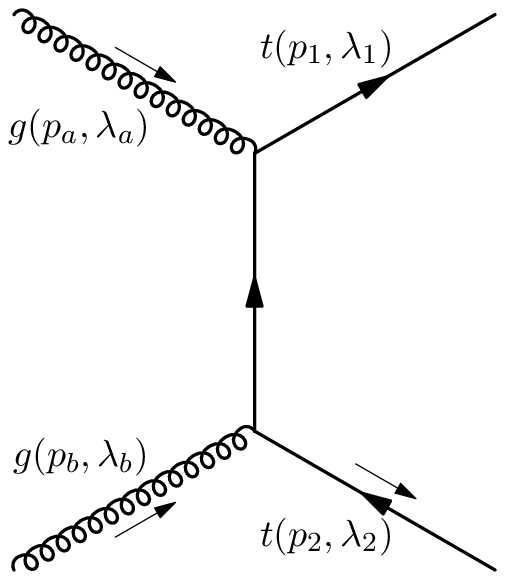}{%
    \texttt{Dgg\_tF\_tree}*\\
    $\Fermion = t$
}\par}
%
%
\subsection{Self-Energy Diagrams}
%
For our calculation we only need to consider self-energy corrections for
fermions and gluons. The SUSY self-energy corrections to fermion propagators
only come from scalar particles:
\begin{equation}\label{eq:pro:seF}
  \mathalign{\input{pro-seF-SF.tex}}
\end{equation}
The $s$-channel gluon propagator gets SUSY corrections from fermion and scalar
loops:
\begin{equation}\label{eq:pro:seG}
  \mathalign{\input{pro-seG-FF.tex}}
  \quad,\quad
  \mathalign{\input{pro-seG-SS.tex}}
  \quad,\quad
  \mathalign{\input{pro-seG-S.tex}}
\end{equation}
The two scalar self-energy diagrams always appear as a pair with the same
coefficient and the same scalar particle in the loop. By inserting these
self-energy corrections in individual lines of the tree level diagrams we obtain
the following \emph{self-energy diagrams}:
\newpage
{\centering
  $\mathalign{\input{pro-qqbar-sG-xseq-1.tex}}
  \quad+\quad
  \mathalign{\input{pro-qqbar-sG-xseq-2.tex}}$
  \par
  \texttt{Dqqbar\_sV\_xseSq}\\
  $ (\Quark,\Fermion,\Scalar)
   =(u^I,\chi^0_j,U_i),\ (d^I,\chi^0_j,D_i),\ (u^I,\chi_j,D_i),\ 
    (d^I,\chi^\cconj_j,U_i),\ (u^I,\Lambda,U_i),\ (d^I,\Lambda,D_i)$
  \par
}
\par\bigskip
{\centering
  $\mathalign{\input{pro-qqbar-sG-xset-1.tex}}
  \quad+\quad
  \mathalign{\input{pro-qqbar-sG-xset-2.tex}}$
  \par
  \texttt{Dqqbar\_sV\_xseSt}\\
  $ (\Fermion,\Scalar)
   =(\chi^0_j,U_i),\ (\chi_j,D_i),\ (t,H^0_i),\ (t,A^0_1),\ (b,H_1),\
    (\Lambda,U_i)$
\par
}
\par\bigskip
{\centering\includegraphics{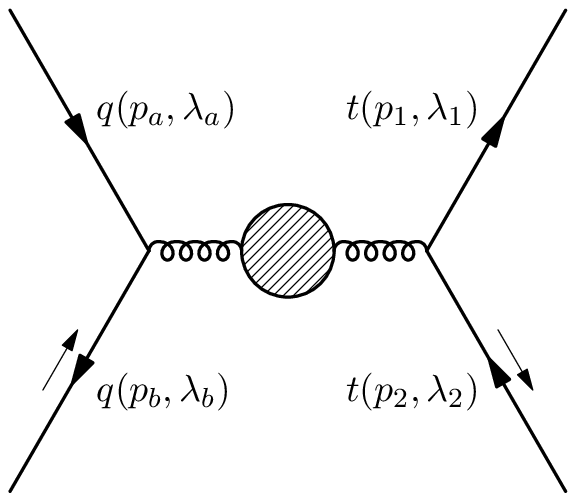}\par
  \texttt{Dqqbar\_sG\_iseF}, \texttt{Dqqbar\_sG\_iseS}\\
  $\Quark = u^I,d^I\quad,\quad
   \Fermion=u^I,t,d^I,b,\Lambda\quad,\quad
   \Scalar=U_i,D_i$
  \par
}
\newpage

{\centering
  $\mathalign{\input{pro-gg-sG-xseg-1.tex}}
  \quad+\quad
  \mathalign{\input{pro-gg-sG-xseg-2.tex}}$
  \par
  \texttt{Dgg\_sG\_xseFg}, \texttt{Dgg\_sG\_xseSg}\\
  $\Fermion=\Lambda\quad,\quad\Scalar=U_i,D_i$
   \par
}
\par\bigskip
{\centering
  $\mathalign{\input{pro-gg-sG-xset-1.tex}}
  \quad+\quad
  \mathalign{\input{pro-gg-sG-xset-2.tex}}$
  \par
  \texttt{Dgg\_sG\_xseSt}\\
  $ (\Fermion,\Scalar)
   =(\chi^0_j,U_i),\ (\chi_j,D_i),\ (t,H^0_i),\ (t,A^0_1),\ (b,H_1),\
    (\Lambda,U_i)$
   \par
}
\par\bigskip
{\centering\includegraphics{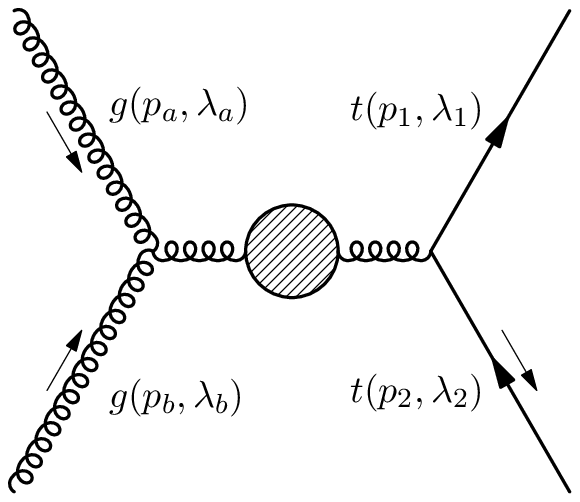}\par
  \texttt{Dgg\_sG\_iseF}, \texttt{gg\_sG\_iseS}\\
  $\Fermion=\Lambda\quad,\quad
   \Scalar=U_i,D_i$
  \par
}
\newpage

{\centering
  $\mathalign{\input{pro-gg-tF-xset-1.tex}}
  \quad+\quad
  \mathalign{\input{pro-gg-tF-xset-2.tex}}$
  \par
  \texttt{Dgg\_tF\_xseSt}*\\
  $ (\Fermion,\Scalar)
   =(\chi^0_j,U_i),\ (\chi_j,D_i),\ (t,H^0_i),\ (t,A^0_1),\ (b,H_1),\
    (\Lambda,U_i)$
  \par
}
\par\bigskip
{\centering
  $\mathalign{\input{pro-gg-tF-xseg-1.tex}}
  \quad+\quad
  \mathalign{\input{pro-gg-tF-xseg-2.tex}}$
  \par
  \texttt{Dgg\_tF\_xseSt}*\\
  $\Fermion=\Lambda\quad,\quad\Scalar=U_i,D_i$
  \par
}
\newpage

{\centering\includegraphics{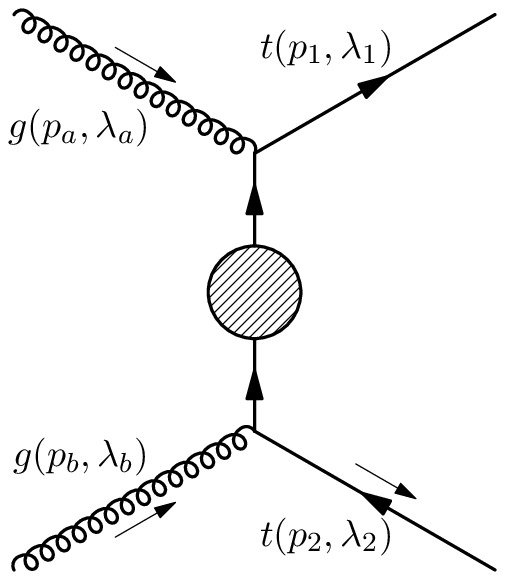}\par
  \texttt{Dgg\_tF\_iseS}*\\
  $ (\Fermion,\Scalar)
   =(\chi^0_j,U_i),\ (\chi_j,D_i),\ (t,H^0_i),\ (t,A^0_1),\ (b,H_1),\
    (\Lambda,U_i)$
  \par
}

In each diagram the hatched blob stands for one of the self-energy corrections
from \eqref{eq:pro:seF} or \eqref{eq:pro:seG}.
\newpage
%
%
\subsection{Vertex Corrections}
%
The prototype vertex corrections for the $q\bar q\to t\bar t$ amplitude are:
\par\bigskip
\noindent
\includediagrams{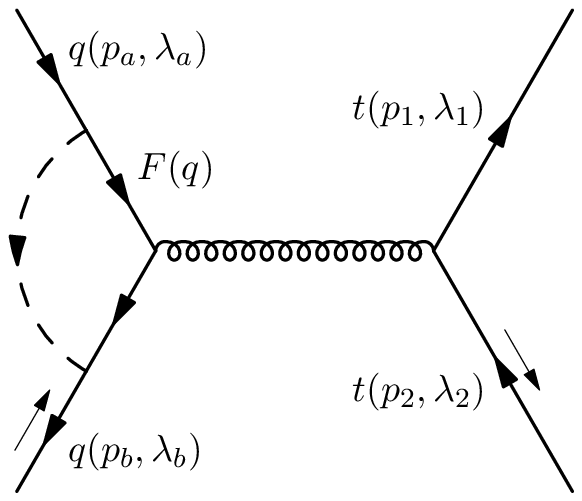}{%
  \texttt{Dqqbar\_sG\_vertSq}\\
  $(\Quark,\Fermion,\Scalar)=(u^I,\Lambda,U_i),\ (d^I,\Lambda,D_i)$%
}%
\includediagrams{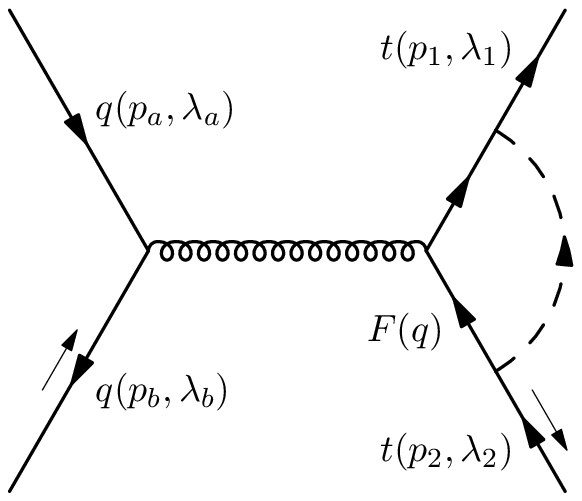}{%
  \texttt{Dqqbar\_sV\_vertSt}\\
  $\Quark=u^I,d^I$\\
  $(\Fermion,\Scalar)=(\Lambda,U_i),\ (t,H^0_1),\ (t,A^0_1),\ (b,H_1)$
}%
\par\bigskip
\noindent\includediagrams{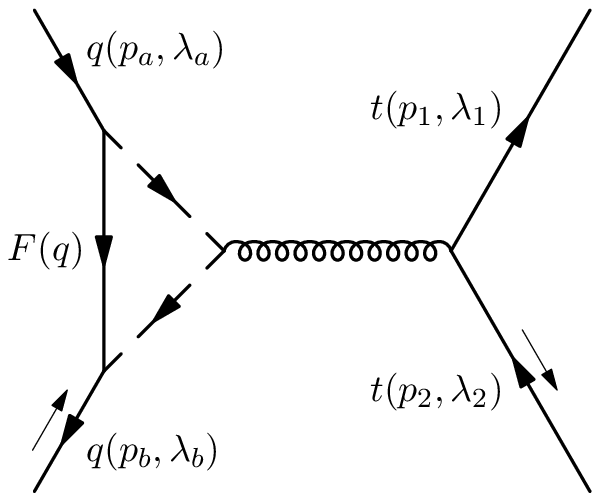}{%
  \texttt{Dqqbar\_sG\_vertSSq}\\
  $ (\Quark,\Fermion,\Scalar)
   =(u^I,\Lambda,U_i),\ (u^I,\chi^0_j,U_i),\ (u^I,\chi_j,D_i),$\\
  $ (d^I,\Lambda,D_i),\ (d^I,\chi^0_j,D_i),\ (d^I,\chi_j,U_i)$
}%
\includediagrams{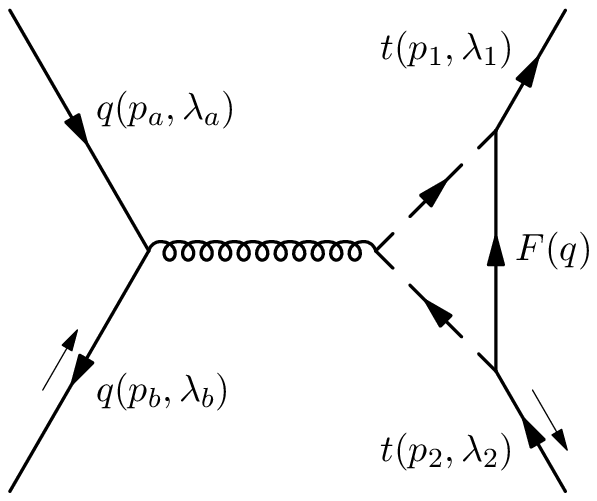}{%
  \texttt{Dqqbar\_sG\_vertSSt}\\
  $\Quark=u^I,d^I$\\
  $(\Scalar,\Fermion)=(\Lambda,U_i),\ (\chi^0_j,U_i),\ (\chi_j,D_i)$
}
\newpage

For the $gg\to t\bar t$ amplitude we distinguish vertex corrections for $s$ and
$t$-channel diagrams. The corrections to the $s$-channel diagrams are:
\par\bigskip
\noindent\includediagrams{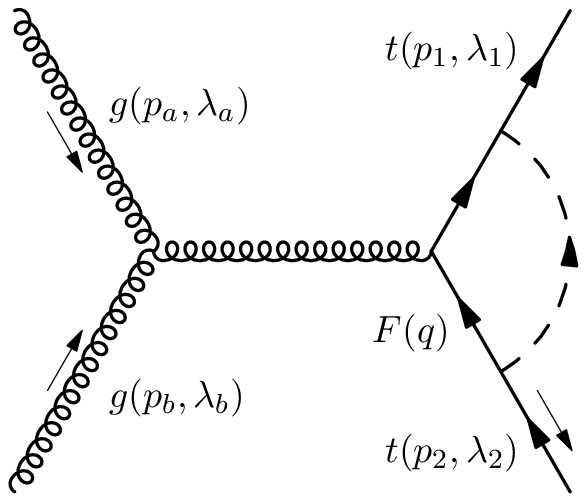}{%
  \texttt{Dgg\_sG\_vertSt}\\
  $(\Fermion,\Scalar)=(\Lambda,U_i),\ (t,H^0_1),\ (t,A^0_1),\ (b,H_1)$
}%
\includediagrams{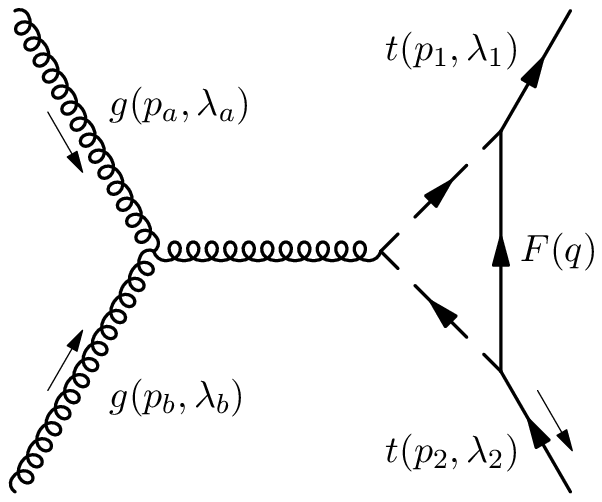}{%
  \texttt{Dgg\_sG\_vertSSt}\\
  $(\Scalar,\Fermion)=(U_i,\Lambda),\ (U_i,\chi^0_j),\ (D_i,\chi_j)$
}
\par\bigskip
{\centering
  $\mathalign{\input{pro-gg-sG-vertFg-ccw.tex}}
  \quad-\quad
  \mathalign{\input{pro-gg-sG-vertFg-cw.tex}}$
  \par\medskip
  \texttt{Dgg\_sG\_vertFg}\\
  $\Fermion=\Lambda$
  \par
}
\par\bigskip
{\centering
  $\mathalign{\input{pro-gg-sG-vertSg-ccw.tex}}
  \quad-\quad
  \mathalign{\input{pro-gg-sG-vertSg-cw.tex}}$
  \par\medskip
  \texttt{Dgg\_sG\_vertSg}\\
  $\Scalar=U_i,D_i$
  \par
}
\newpage

{\centering
  $\mathalign{\input{pro-gg-sS-vertFg-ccw.tex}}
  \quad+\quad
  \mathalign{\input{pro-gg-sS-vertFg-cw.tex}}$
  \par\medskip
  \texttt{Dgg\_sS\_vertFg}\\
  $\Fermion=t,b\quad,\quad\Scalar=H^0_1,A^0_1$
  \par
}
{\centering
  $\mathalign{\input{pro-gg-sS-vertSg-ccw.tex}}
  \quad+\quad
  \mathalign{\input{pro-gg-sS-vertSg-cw.tex}}$
  \par\medskip
  \texttt{Dgg\_sS\_vertSg}\\
  $\Scalar_1=U_i,D_i\quad,\quad\Scalar_2=H^0_1,A^0_1$
  \par
}
\par\bigskip

\hfillstar
\includediagrams{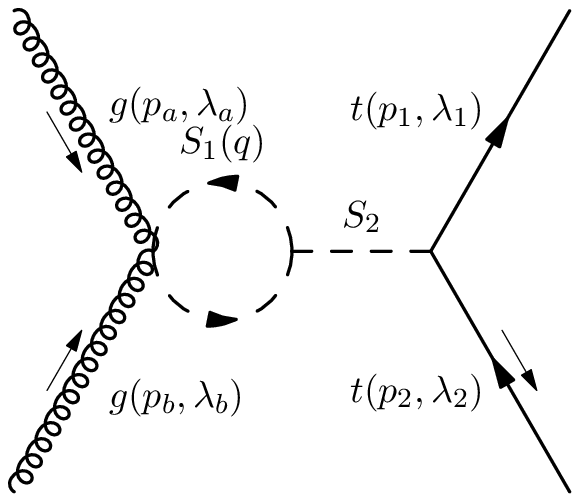}{%
  \texttt{Dgg\_sS\_vertSSg}\\
  $\Scalar_1=U_i,D_i\quad,\quad\Scalar_2=H^0_1,A^0_1$
  \par
}\hfillstar
\newpage

The $t$-channel vertex corrections are:
\par\bigskip
\noindent\includediagrams{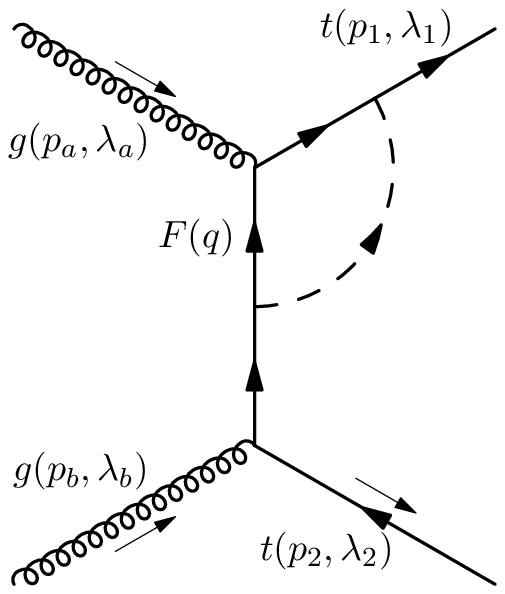}{%
  \texttt{Dgg\_tF\_vertS1}*\\
  $(\Fermion,\Scalar)=(\Lambda,U_i),\ (t,H^0_1),\ (t,A^0_1),\ (b,H_1)$
}%
\includediagrams{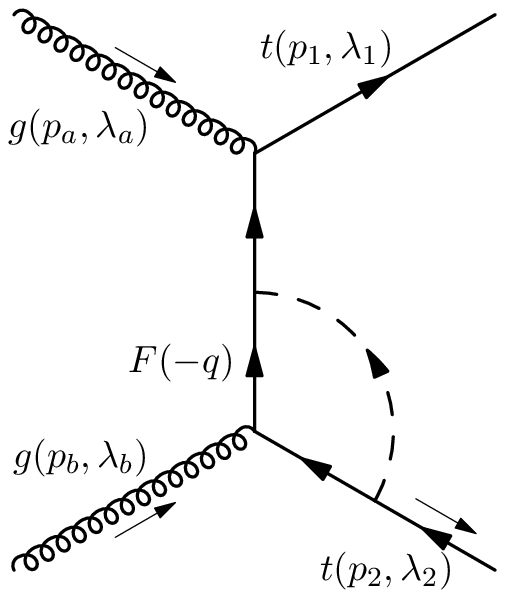}{%
  \texttt{Dgg\_tF\_vertS2}*\\
  $(\Fermion,\Scalar)=(\Lambda,U_i),\ (t,H^0_1),\ (t,A^0_1),\ (b,H_1)$
}
\par\bigskip
\noindent\includediagrams{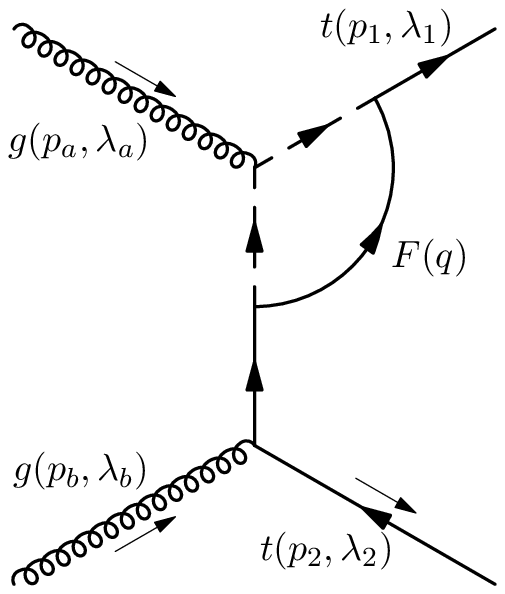}{%
  \texttt{Dgg\_tF\_vertSS1}*\\
  $(\Scalar,\Fermion)=(U_i,\Lambda),\ (U_i,\chi^0_j),\ (D_i,\chi_j)$
}%
\includediagrams{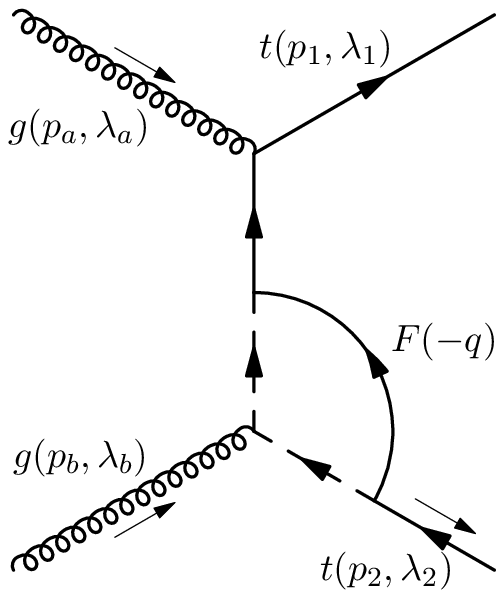}{%
  \texttt{Dgg\_tF\_vertSS2}*\\
  $(\Scalar,\Fermion)=(U_i,\Lambda)$
}
\newpage
%
%
\subsection{Box Diagrams}
%
\label{sec:boxes}
{\centering\includegraphics{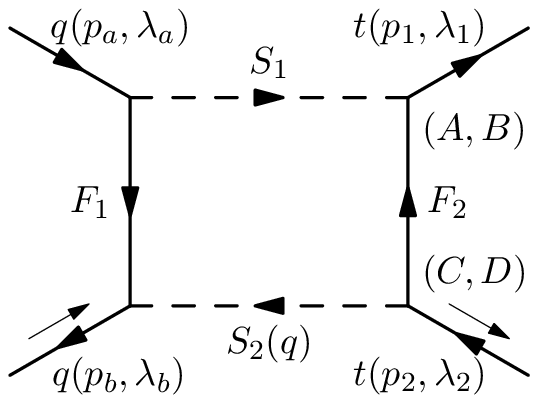}\\
  \texttt{Dqqbar\_boxSS}\\
  $ (\Quark,\Scalar_1,\Scalar_2,\Fermion_1,\Fermion_2)
   =(u^I,U_i,U_j,\Lambda,\Lambda),\
    (u^I,U_i,U_j,\Lambda,\chi^0_k),\ (u^I,U_i,U_j,\chi^0_k,\Lambda),$
  \\
  $ (d^I,D_i,D_j,\Lambda,\chi_k),\ (d^I,U_i,U_j,\chi_k,\Lambda)$
  \par
}%
\par\bigskip
\noindent\includediagrams{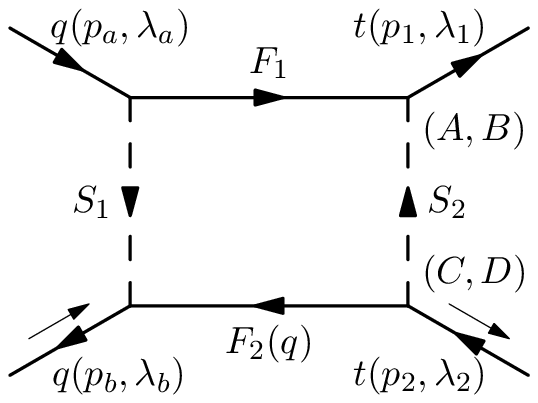}{%
  \texttt{Dqqbar\_fboxSS}\\
  $ (\Quark,\Fermion_1,\Fermion_2,\Scalar_1,\Scalar_2)
   =(u^I,\Lambda,\Lambda,U_i,U_j),$\\
  $ (d^I,\Lambda,\Lambda,D_i,U_j),\ (u^I,\chi^0_k,\Lambda,U_i,U_j),$\\
  $ (u^I,\Lambda,\chi^0_k,U_i,U_j),\ (d^I,\chi^0_k,\Lambda,D_i,U_j),$\\
  $ (d^I,\Lambda,\chi^0_k,D_i,U_j)$
}%
\includediagrams{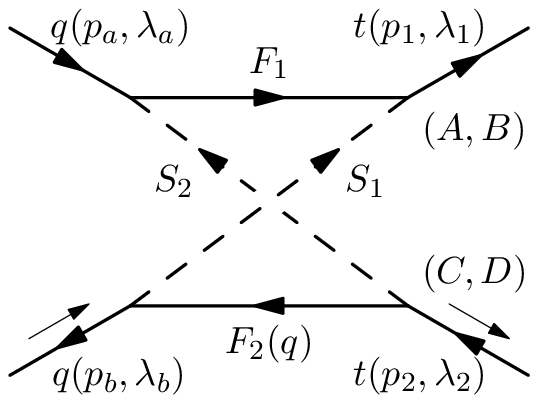}{%
  \texttt{Dqqbar\_fboxSSx}**\\
  $ (\Quark,\Fermion_1,\Fermion_2,\Scalar_1,\Scalar_2)
   =(u^I,\Lambda,\Lambda,U_i,U_j),$\\
  $ (d^I,\Lambda,\Lambda,D_i,U_j),\ (u^I,\chi^0_k,\Lambda,U_i,U_j),$\\
  $ (u^I,\Lambda,\chi^0_k,U_i,U_j),\ (d^I,\chi^0_k,\Lambda,D_i,U_j),$\\
  $ (d^I,\Lambda,\chi^0_k,D_i,U_j)$
}%
\par\bigskip
\noindent\includediagrams{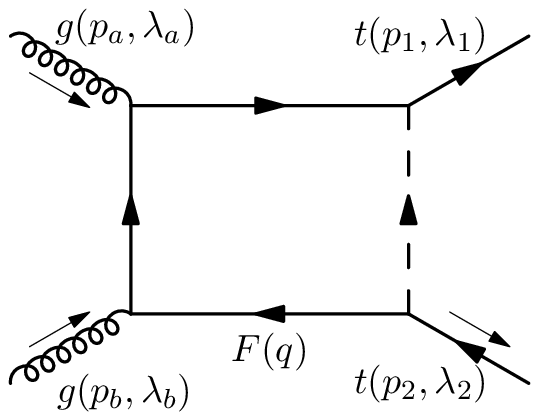}{
  \texttt{Dgg\_boxFS}*\\
  $(\Fermion,\Scalar)=(\Lambda,U_i),\ (t,H^0_1),\ (t,A^0_1),\ (b,H_i)$
}%
\includediagrams{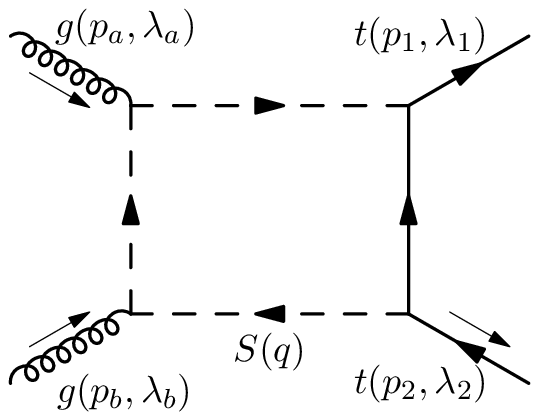}{%
  \texttt{Dgg\_boxSF}*\\
  $(\Scalar,\Fermion)=(U_i,\Lambda),\ (U_i,\chi^0_j),\ (D_i,\chi_j)$
}%
\newpage

{\centering\includegraphics{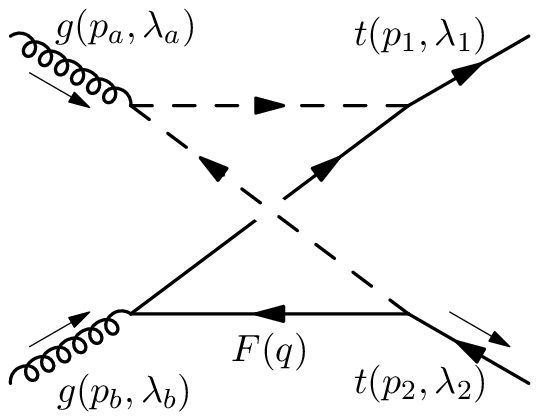}\\
  \texttt{Dgg\_boxSFx}*\\
  $(\Scalar,\Fermion)=(U_i,\Lambda)$
  \par
}

%% file: pro-seF-SF.tex
\raisebox{-17.2591bp}{\includegraphics{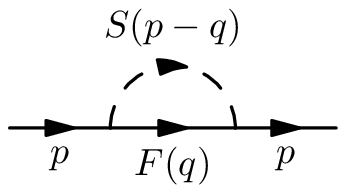}}

%% file: pro-seG-FF.tex
\raisebox{-35.2591bp}{\includegraphics{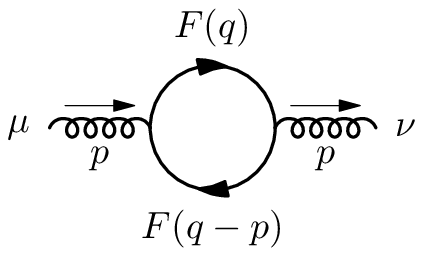}}

%% file: pro-seG-SS.tex
\raisebox{-35.2591bp}{\includegraphics{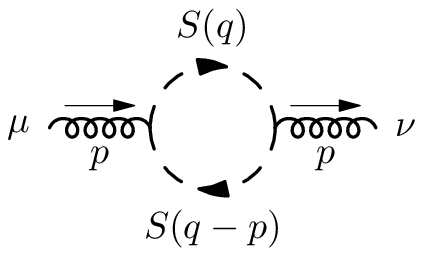}}

%% file: pro-seG-S.tex
\raisebox{-13.1682bp}{\includegraphics{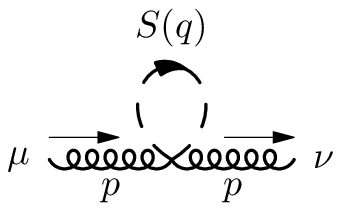}}

%% file: pro-qqbar-sG-xseq-1.tex
\raisebox{-69.682bp}{\includegraphics{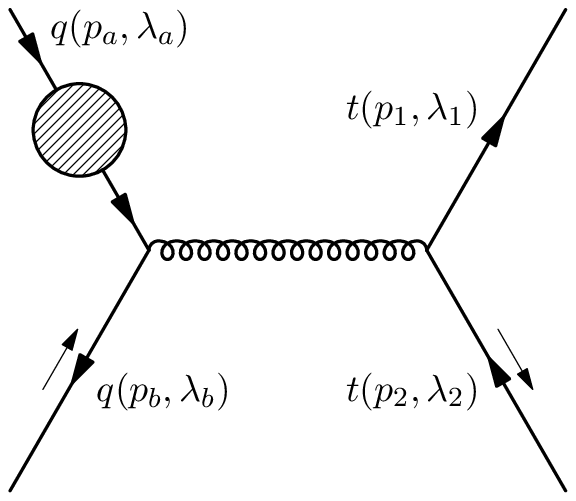}}

%% file: pro-qqbar-sG-xseq-2.tex
\raisebox{-69.9387bp}{\includegraphics{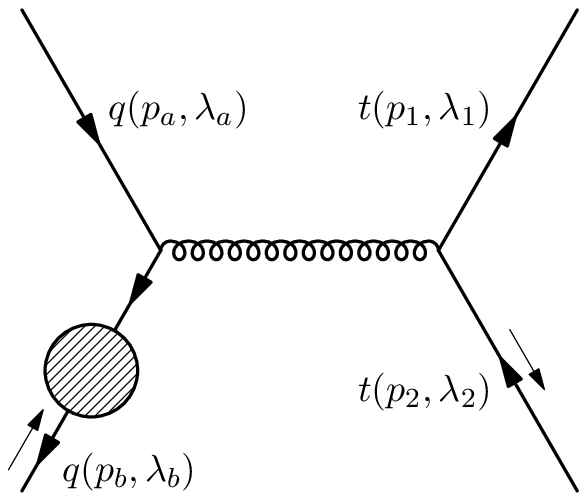}}

%% file: pro-qqbar-sG-xset-1.tex
\raisebox{-69.682bp}{\includegraphics{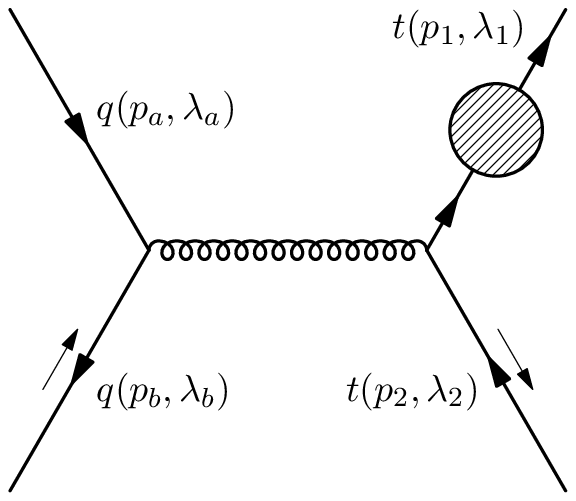}}

%% file: pro-qqbar-sG-xset-2.tex
\raisebox{-69.9387bp}{\includegraphics{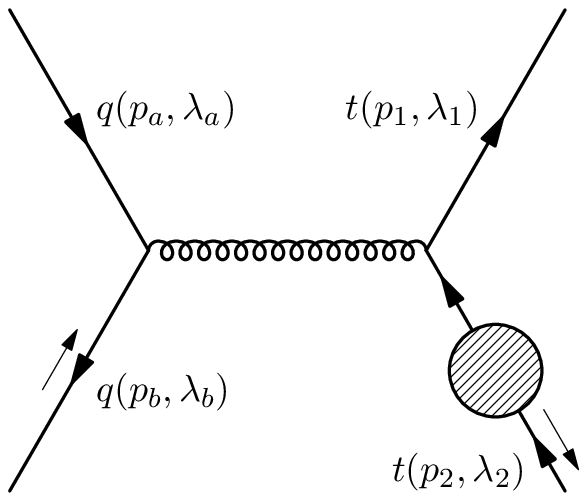}}

%% file: pro-gg-sG-xseg-1.tex
\raisebox{-69.682bp}{\includegraphics{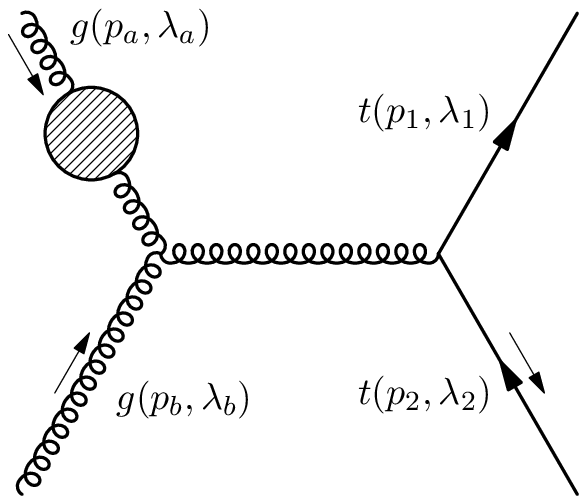}}

%% file: pro-gg-sG-xseg-2.tex
\raisebox{-71.3137bp}{\includegraphics{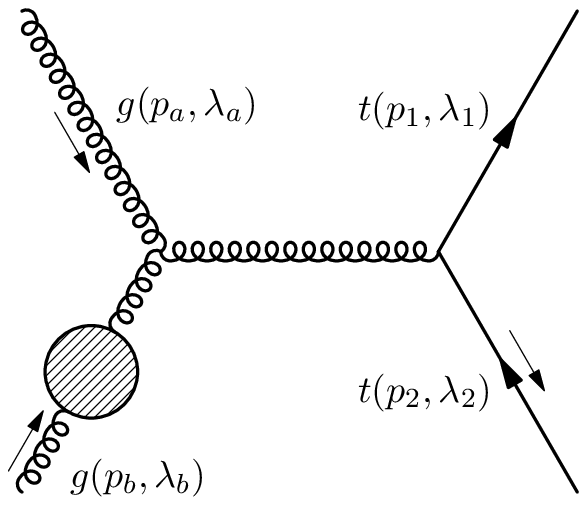}}

%% file: pro-gg-sG-xset-1.tex
\raisebox{-69.682bp}{\includegraphics{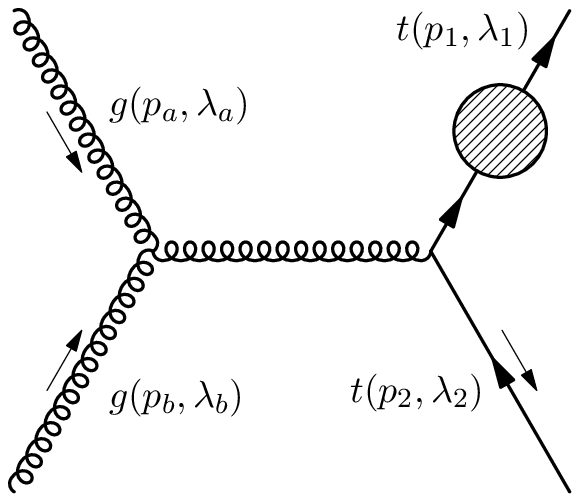}}

%% file: pro-gg-sG-xset-2.tex
\raisebox{-69.9387bp}{\includegraphics{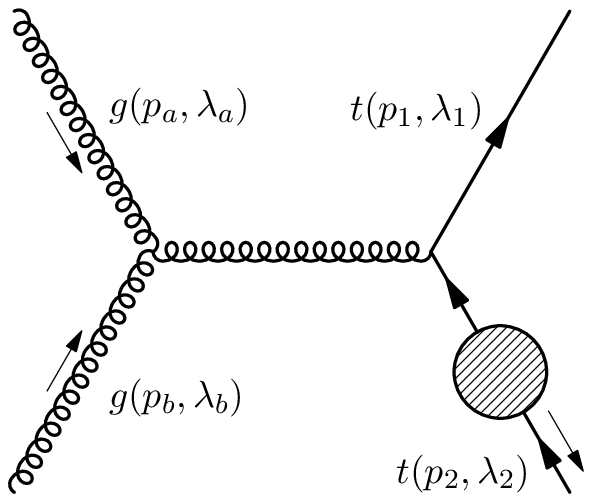}}

%% file: pro-gg-tF-xset-1.tex
\raisebox{-80.4bp}{\includegraphics{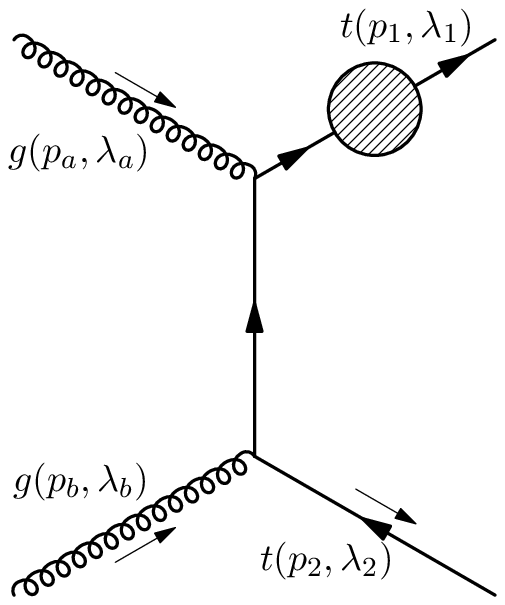}}

%% file: pro-gg-tF-xset-2.tex
\raisebox{-89.8556bp}{\includegraphics{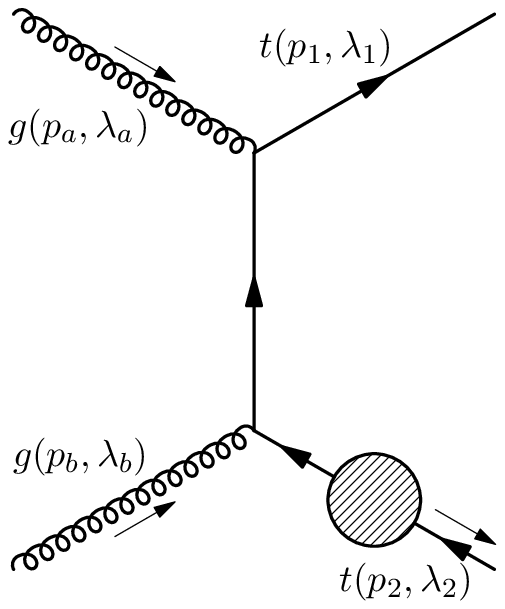}}

%% file: pro-gg-tF-xseg-1.tex
\raisebox{-80.4bp}{\includegraphics{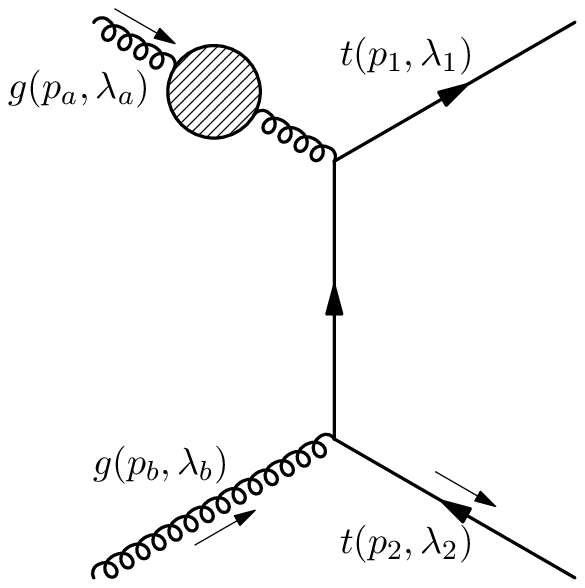}}

%% file: pro-gg-tF-xseg-2.tex
\raisebox{-85.0759bp}{\includegraphics{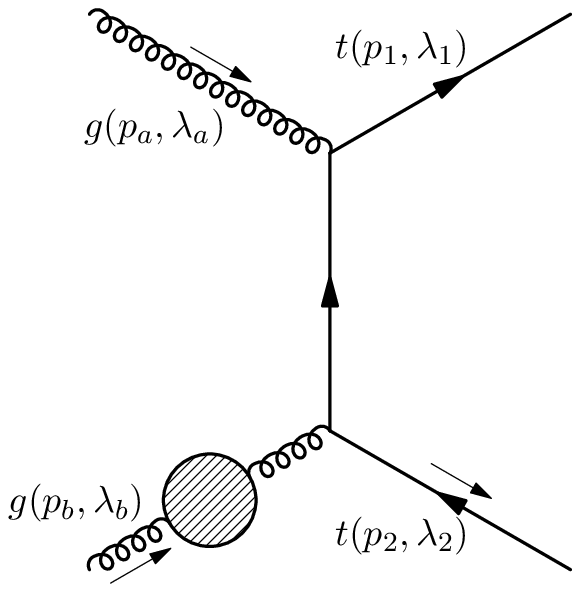}}

%% file: pro-gg-sG-vertFg-ccw.tex
\raisebox{-69.682bp}{\includegraphics{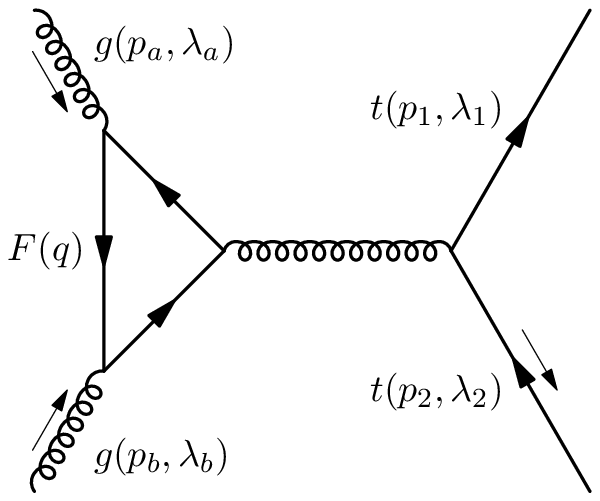}}

%% file: pro-gg-sG-vertFg-cw.tex
\raisebox{-69.682bp}{\includegraphics{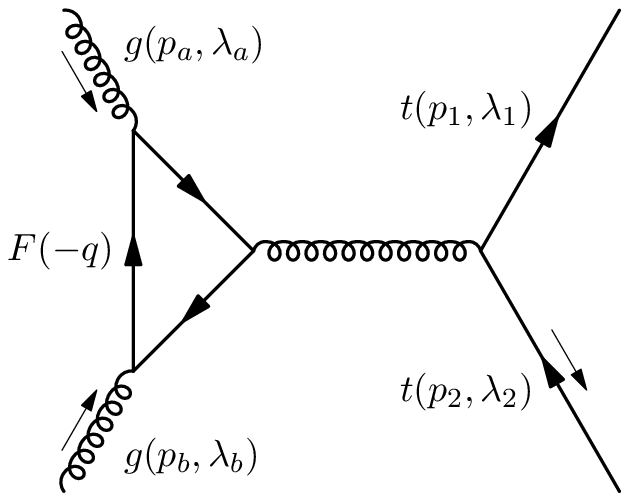}}

%% file: pro-gg-sG-vertSg-ccw.tex
\raisebox{-69.682bp}{\includegraphics{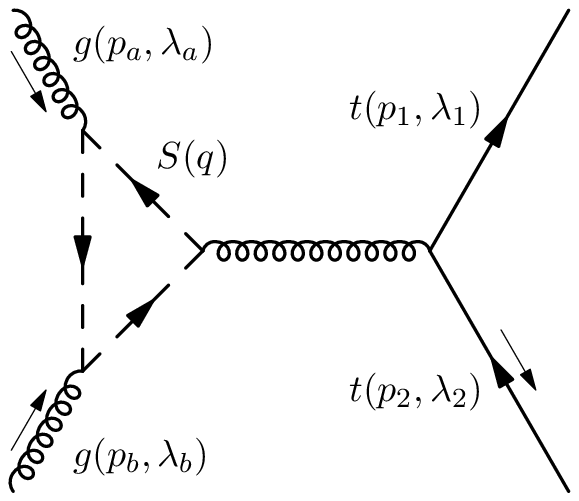}}

%% file: pro-gg-sG-vertSg-cw.tex
\raisebox{-69.682bp}{\includegraphics{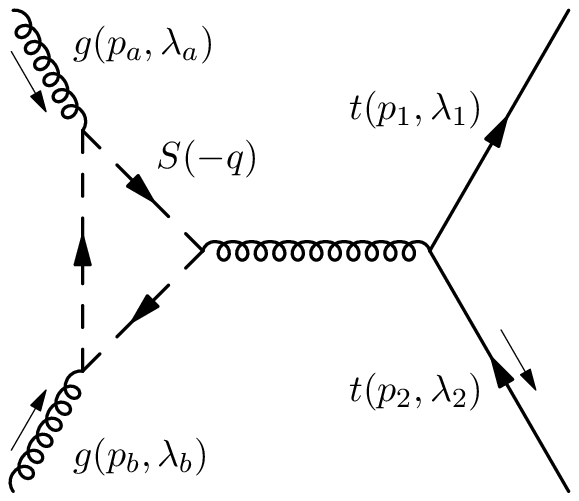}}

%% file: pro-gg-sS-vertFg-ccw.tex
\raisebox{-69.682bp}{\includegraphics{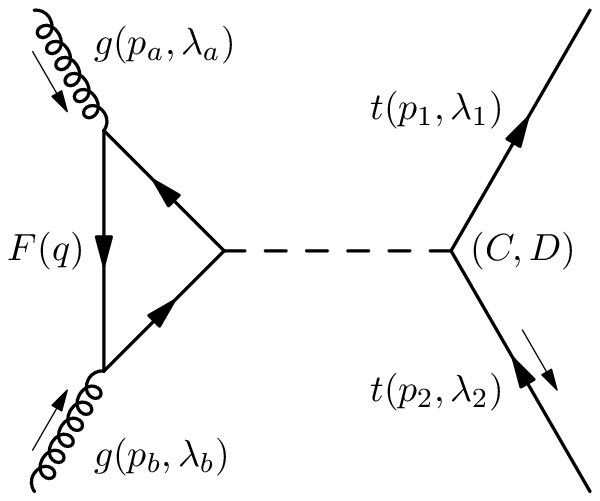}}

%% file: pro-gg-sS-vertFg-cw.tex
\raisebox{-69.682bp}{\includegraphics{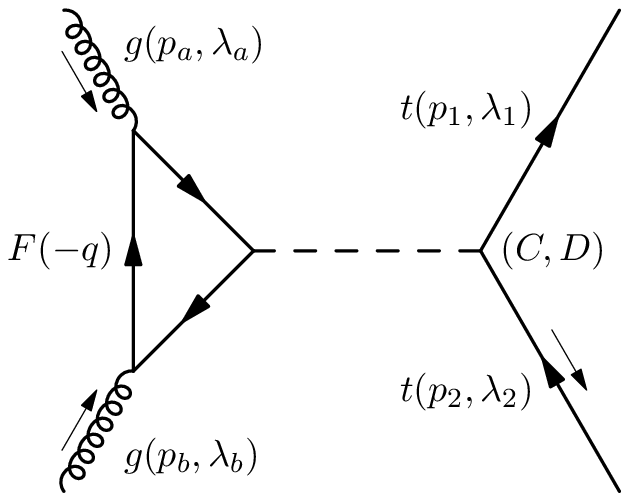}}

%% file: pro-gg-sS-vertSg-ccw.tex
\raisebox{-69.682bp}{\includegraphics{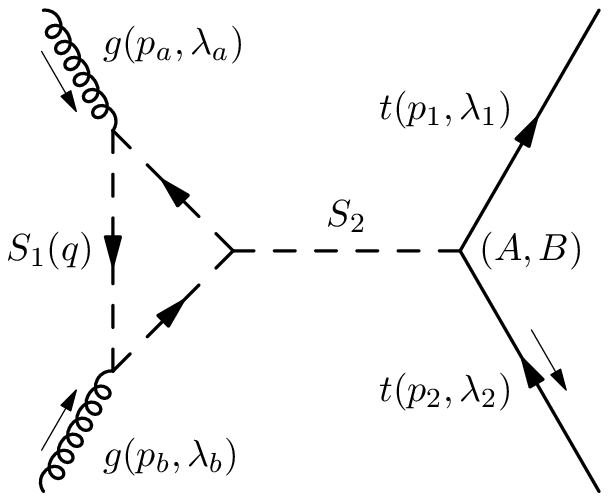}}

%% file: pro-gg-sS-vertSg-cw.tex
\raisebox{-69.682bp}{\includegraphics{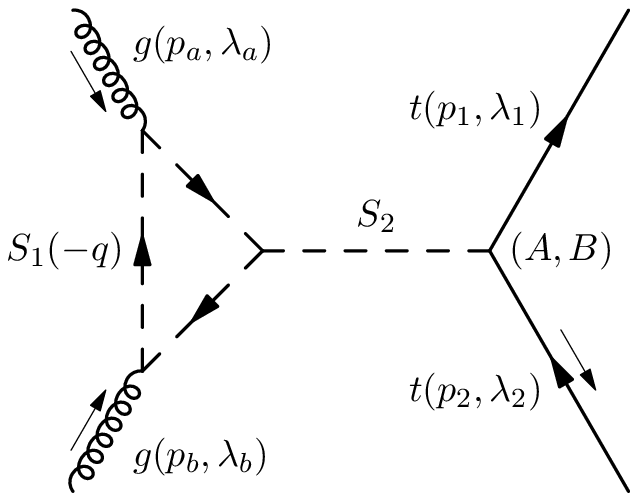}}

%% file: results.tex
\section{Cross Sections and Asymmetries}
\label{sec:results}
In this section we present our results for the SUSY corrections to polarised
$t\bar t$ production cross sections, which we calculated for each of the 10
Snowmass benchmarks detailed in \cite{Snowmass02}. To calculate the masses of
the supersymmetric particles and run the couplings to the TeV scale we used the
program \texttt{SOFTSUSY} by B.\ C.\ Allanach \cite{Softsusy}. The decay widths
of the MSSM Higgs particles were calculated with the program \texttt{HDECAY} by
Djouadi, Kalinowski and Spira \cite{Djouadi97}. The Feynman rules for the MSSM
vertices were taken from J. Rosiek's paper \cite{Rosiek95}. We compare our
parton level cross sections with the results obtained in the leading log
approximation \cite{Beccaria04}. Then we discuss our results for the total
$pp\to t\bar t$ cross section and the double helicity asymmetries introduced in
\cite{Bourrely90}.

\begin{figure}
  \centering
  \includegraphics{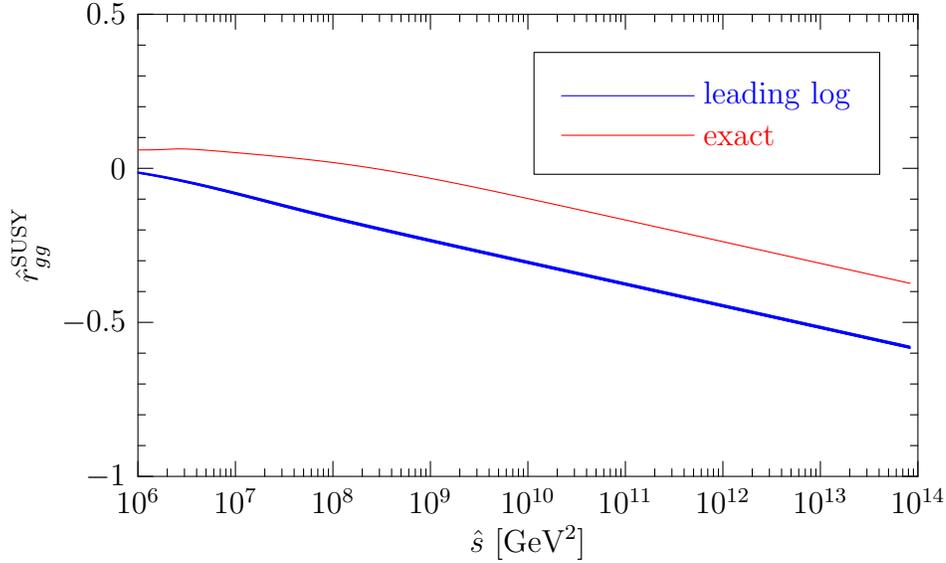}
  \caption{\label{fig:res:ll_gg} SUSY corrections to the $gg\to t\bar t$ cross
    section in the ``exact'' calculation and the leading log approximation for
    benchmark 5 of \cite{Snowmass02}. The width of the leading log graph
    reflects the uncertaincy due to the choice of the universal SUSY scale.}
\end{figure}

\begin{figure}
  \centering
  \includegraphics{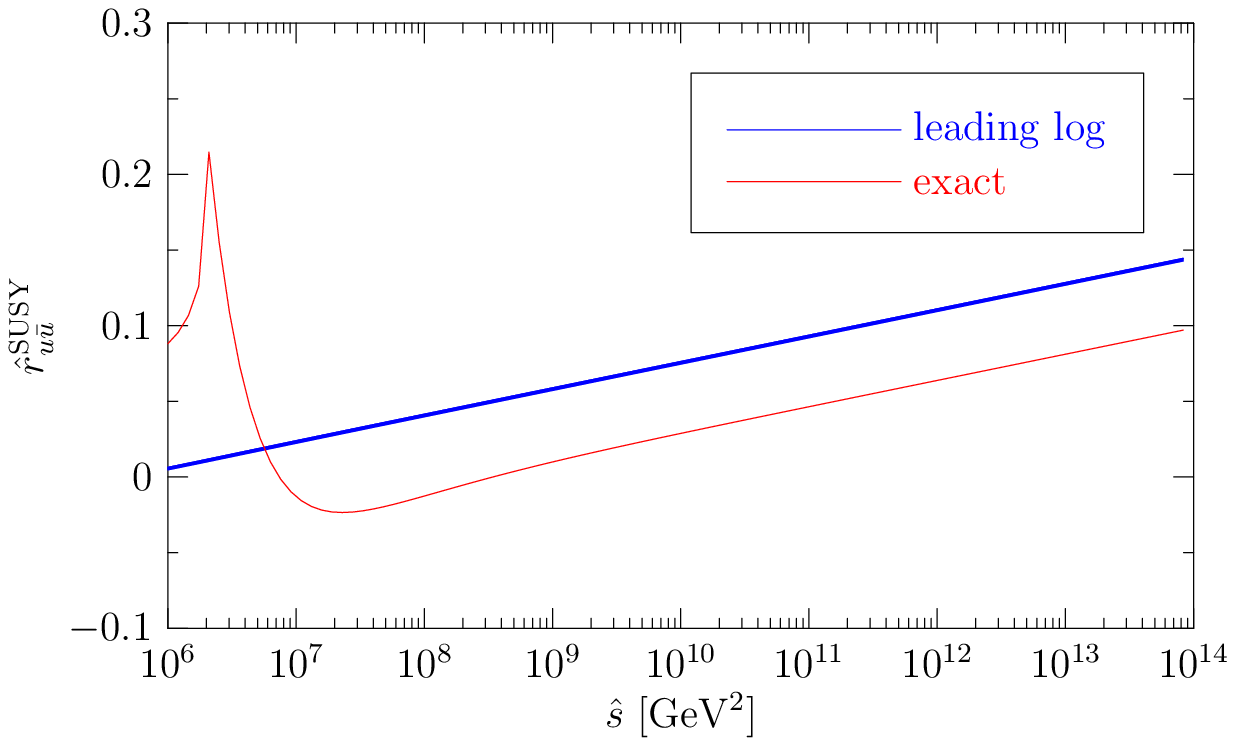}
  \caption{\label{fig:res:ll_uu} SUSY corrections to the $u\bar u\to t\bar t$
    cross section in the ``exact'' calculation and the leading log approximation
    for benchmark 5 of \cite{Snowmass02}. The results for first and second
    generation up-type quarks are identical. The width of the leading log graph
    reflects the uncertainty due to the choice of the universal SUSY scale.}
\end{figure}

\begin{figure}
  \centering
  \includegraphics{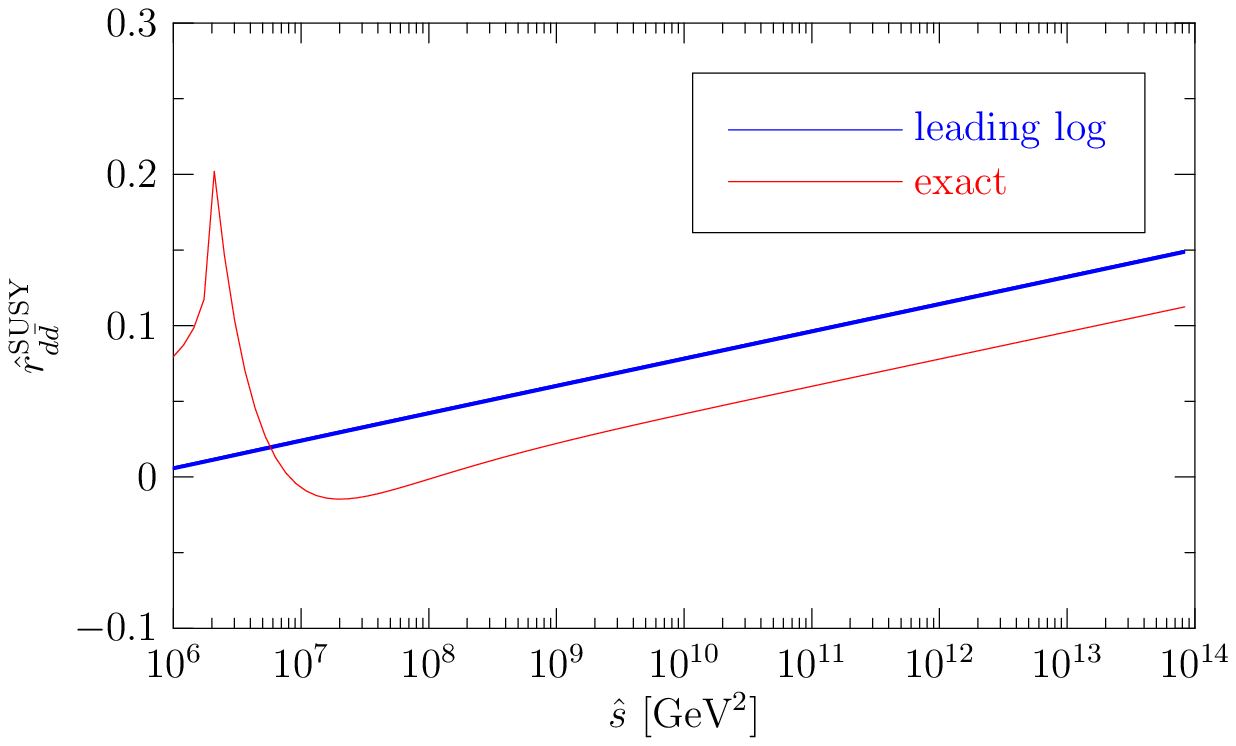}
  \caption{\label{fig:res:ll_dd} SUSY corrections to the $d\bar d\to t\bar t$
    cross section in the ``exact'' calculation and the leading log approximation
    for benchmark 5 of \cite{Snowmass02}. The results for first and second
    generation down-type quarks are identical. The width of the leading log
    graph reflects the uncertainty due to the choice of the universal SUSY
    scale.}
\end{figure}

Let $\hat\sigma_i$ denote the total cross section for the process $i\to t\bar
t$, where the initial state $i$ can be a gluon pair ($gg$), a light up-type
quark-antiquark pair ($u\bar u$) or a light down type quark-antiquark pair
($d\bar d$). We regard $\hat\sigma_i$ as a function of the variable $\hat
s\equiv M_{t\bar t}^2$, where $M_{t\bar t}$ is the invariant mass of the
top-antitop pair. For each of these cross sections we have calculated the
leading order contribution $\hat\sigma_i^{\text{LO}}$ and the SUSY corrections
$\hat\sigma_i^{\text{SUSY}}$ due to the diagrams listed in the previous
section. The SUSY corrections can be split into super-QCD (SQCD) corrections and
super-electroweak (SEW) corrections. The SQCD corrections are of order
$\order(\alpha_s^3)$ and the SEW correction of order $\order(\alpha\alpha_s^2)$.
Consequently the SEW corrections are one order of magnitude smaller than the
SQCD corrections. We also define the ratios
\begin{equation}
    \hat r_i(\hat s)
  = \frac{\hat\sigma_i^{\text{SUSY}}(\hat s)}{\hat\sigma_i^{\text{LO}}(\hat s)}
  \eqpunct.
\end{equation}
\mrefs{Figures \ref{fig:res:ll_gg}, \ref{fig:res:ll_uu} and \ref{fig:res:ll_dd}}
show a comparison of our ``exact'' ratios with the results obtained in the
leading log approximation by Beccaria, Renard and Verzegnassi \cite{Beccaria04}.
We have used Snowmass benchmark 5 to compute the exact cross section, but the
observations stated here are true for any of the 10 Snowmass benchmarks. The
only SUSY inputs in the leading log approximation are $\tan\beta$ and a
universal SUSY mass scale $M_{\text{SUSY}}$. Sensible values for this scale lie
anywhere between the mass of the lightest and the mass of the heaviest SUSY
particle. The widths of the leading log graphs in \mrefs{figures
\ref{fig:res:ll_gg}, \ref{fig:res:ll_uu} and \ref{fig:res:ll_dd}} reflect this
uncertainly. We see that, in the leading log approximation, the ratios $\hat
r_i^{\text{SUSY}}$ are proportional to $\log(\hat s/M_{\text{SUSY}}^2)$. For
$\hat s\gtrsim\unit{10^9}{GeV}$ the exact ratio runs linear with the same slope,
but with a constant offset to the leading log graph. For very large centre of
mass energies this offset becomes negligible. Therefore our results agree with
the leading log approximation in the high energy limit. However, for collision
energies that are achievable at the LHC we see that the leading log
approximation fails to reproduce the exact results of the full SUSY
calculation, even with the offset included.

To obtain the $pp\to t\bar t$ cross sections, the parton level cross sections
$\hat\sigma_i$ were folded with the CTEQ6L1 set of the CTEQ v6.51 parton
distribution functions \cite{CTEQ6}. For the proton-proton collision we assumed
a centre of mass energy of \unit{14}{TeV}.  Let
$d\sigma_{\lambda_1\lambda_2}/dM_{t\bar t}$ denote the invariant mass
differential cross section for producing a top quark with helicity $\lambda_1$
and an anti-top quark with helicity $\lambda_2$. Then we define
\begin{subequations}\label{eq:res:asymmetries}
  \begin{align}
       \frac{d\sigma_{\text{tot}}}{dM_{t\bar t}}
    &=  \frac{d\sigma_{++}}{dM_{t\bar t}}
      + \frac{d\sigma_{--}}{dM_{t\bar t}}
      + \frac{d\sigma_{+-}}{dM_{t\bar t}}
      + \frac{d\sigma_{-+}}{dM_{t\bar t}}
    \eqpunct,\\
       \frac{d\sigma_{LL}}{dM_{t\bar t}}
    &=  \frac{d\sigma_{++}}{dM_{t\bar t}}
      + \frac{d\sigma_{--}}{dM_{t\bar t}}
      - \frac{d\sigma_{+-}}{dM_{t\bar t}}
      - \frac{d\sigma_{-+}}{dM_{t\bar t}}
    \eqpunct,\\
       \frac{d\sigma_{PV}}{dM_{t\bar t}}
    &=  \frac{d\sigma_{+-}}{dM_{t\bar t}}
      - \frac{d\sigma_{-+}}{dM_{t\bar t}}
    \eqpunct.
  \end{align}
\end{subequations}
For each combination we indicate the leading order and SUSY contributions by
superscripts `LO' and `SUSY', respectively. The parity even combinations
$d\sigma^{\text{SUSY}}_{\text{tot}}$ and $d\sigma^{\text{SUSY}}_{LL}$ are
dominated by the SQCD corrections.  However, for the parity odd combination
$d\sigma^{\text{SUSY}}_{PV}$ the SQCD corrections are zero, since parity is
conserved in super-QCD. For the asymmetries and the SUSY corrections we define
the ratios
\begin{equation}\label{eq:res:ratios}
   r^{\text{LO}}_{LL/PV}(M_{t\bar t})
  =\frac{d\sigma^{\text{LO}}_{LL/PV}/dM_{t\bar t}}
        {d\sigma^{\text{LO}}_{\text{tot}}/dM_{t\bar t}}
  \qquad,\qquad
   r^{\text{SUSY}}_{\text{tot}/LL/PV}(M_{t\bar t})
  =\frac{d\sigma^{\text{SUSY}}_{\text{tot}/LL/PV}/dM_{t\bar t}}
        {d\sigma^{\text{LO}}_{\text{tot}}/dM_{t\bar t}}
  \eqpunct.
\end{equation}

\begin{figure}
  \centering
  \includegraphics{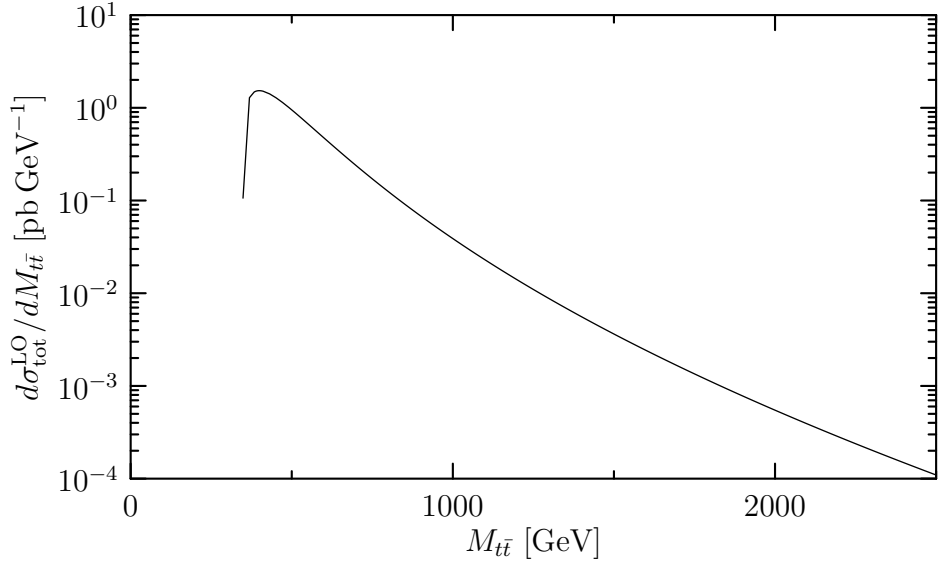}
  \caption{\label{fig:res:tree_tot_mtt} Leading order results for the invariant
  mass differential cross section.}
\end{figure}
\begin{figure}
  \centering
  \includegraphics{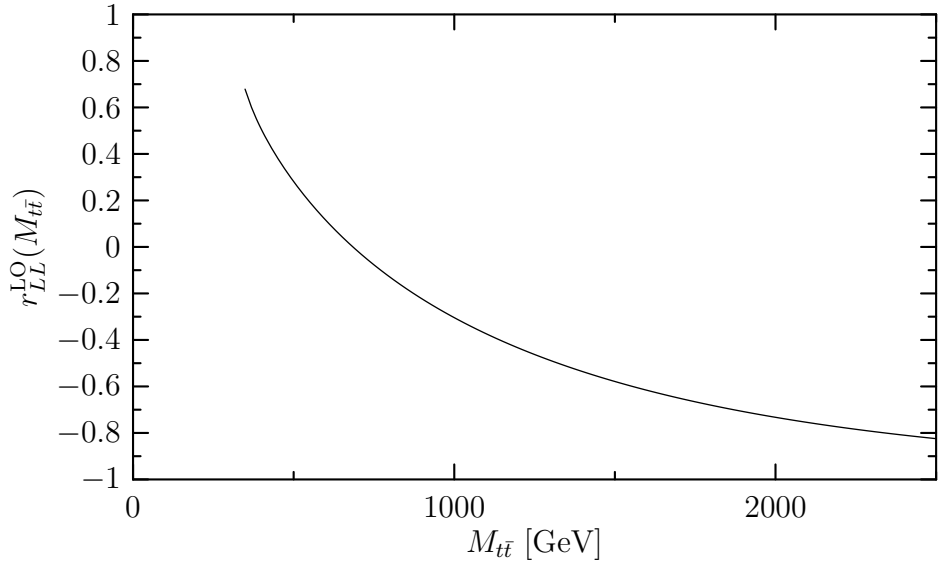}
  \caption{\label{fig:res:tree_al_mtt} Leading order results for the invariant
  mass $LL$ asymmetry.}
\end{figure}
\mrefs{Figures \ref{fig:res:tree_tot_mtt} and \ref{fig:res:tree_al_mtt}} show
the results for $d\sigma^{\text{LO}}_{\text{tot}}/dM_{t\bar t}$ and
$r^{\text{LO}}_{LL}(M_{t\bar t})$, respectively.
Since there is no parity violation at leading order the ratio
$r^{\text{LO}}_{PV}$ is identically zero.

\begin{figure}
  \centering
  \includegraphics{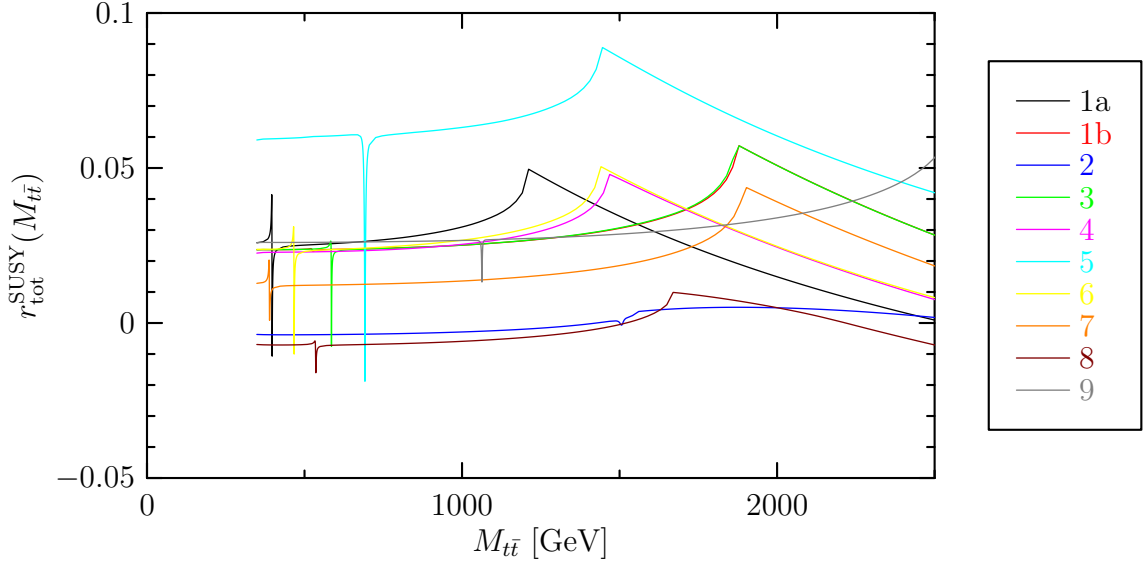}
  \caption{\label{fig:res:tot_mtt} SUSY corrections to the invariant mass
  differential cross section for the Snowmass benchmarks. The numbers in the
  legend refer to the labelling of the benchmarks in \cite{Snowmass02}.}
\end{figure}

\begin{figure}
  \centering
  \includegraphics{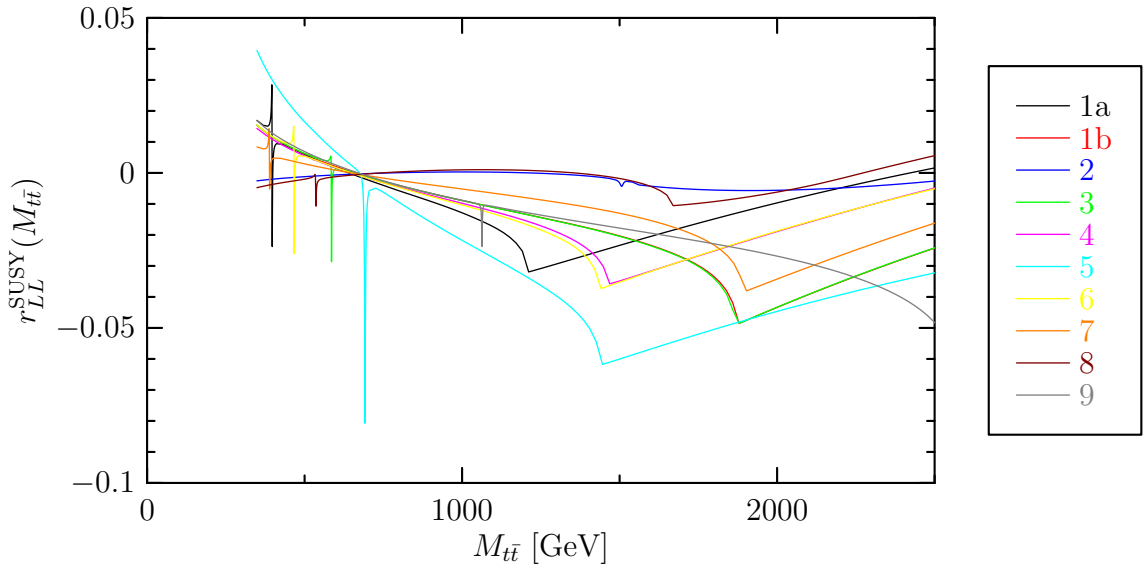}
  \caption{\label{fig:res:al_mtt} SUSY corrections to the invariant mass $LL$
  asymmetry for the Snowmass benchmarks. The numbers in the legend refer to the
  labelling of the benchmarks in \cite{Snowmass02}.}
\end{figure}

\mrefs{Figures \ref{fig:res:tot_mtt} and \ref{fig:res:al_mtt}} show the ratios
$r^{\text{SUSY}}_{\text{tot}}(M_{t\bar t})$ and $r^{\text{SUSY}}_{LL}(M_{t\bar
t})$ for each of the 10 Snowmass benchmarks.
We see that the SUSY corrections to the $t\bar t$ cross section can be as large
as 10\% of the leading order cross section, but typically only reach the 5\%
level. In both plots we see ``resonance peaks'' located at the masses of the
heavy and the pseudo-scalar Higgs ($H^0_1$ and $A^0_1$ in the notation of
\cite{Rosiek95}). They come from the scalar $s$-channel propagators in the
diagrams labelled \texttt{Dgg\_sS\_vertFg}, \texttt{Dgg\_sS\_vertSg} and
\texttt{Dgg\_sS\_vertSSg} in \mref{section}{sec:pro}. 
Note that these particles have negligible coupling to the light quarks
extracted from the incoming hadrons, so that they arise from the gluon fusion
contribution, in which the exchanged scalars are connected to the
incoming gluons via a gluino or squark triangle.
Moreover,  due to the
fermion triangle the sign of the
contribution from
 diagram \texttt{Dgg\_sS\_vertFg} is opposite
that of \texttt{Dgg\_sS\_vertSg} and \texttt{Dgg\_sS\_vertSSg}. This explains
why we get downward-pointing resonance peaks for some of the benchmarks. Also
note that, for all 10 benchmarks, the difference of the heavy and the
pseudo-scalar Higgs masses is much smaller than their decay widths. Consequently
we can only see two distinct peaks when these peaks have opposite signs. The
kinks occurring between 1 and \unit{2}{TeV} coincide, for each benchmark, with
twice the gluino mass and can therefore be understood as a threshold effect
due to the box diagrams \texttt{Dqqbar\_fboxSS} and \texttt{Dgg\_boxFS}.

\mref{Figure}{fig:res:pv_mtt} shows the SUSY corrections to the parity violating
asymmetry for each of the 10 benchmarks. 
\begin{figure}
  \centering
  \includegraphics{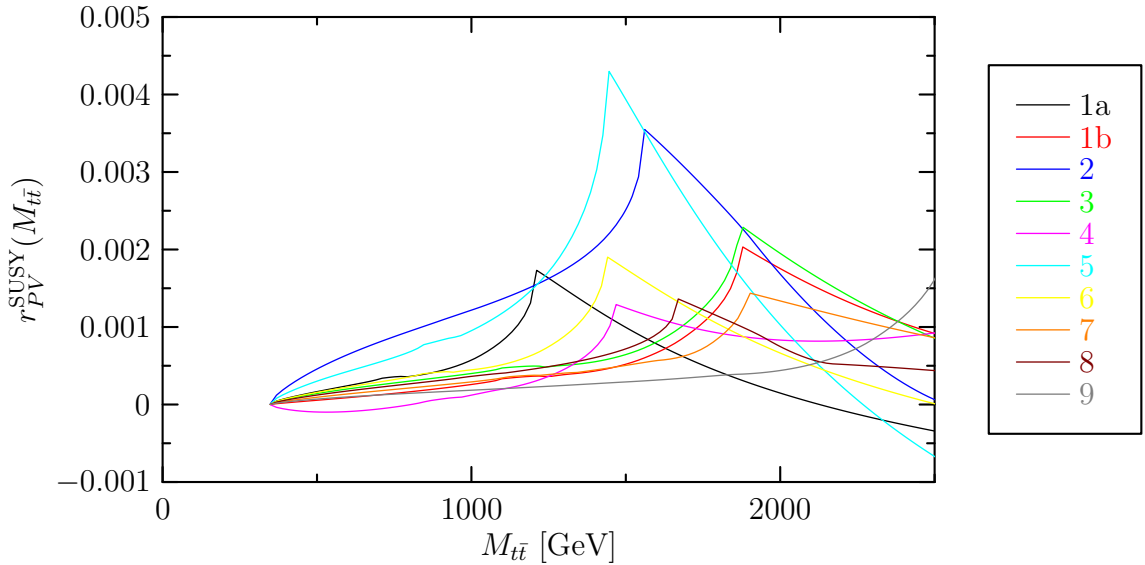}
  \caption{\label{fig:res:pv_mtt} SUSY corrections to the invariant mass $PV$
  asymmetry for the Snowmass benchmarks. The numbers in the legend refer to the
  labelling of the benchmarks in \cite{Snowmass02}.}
\end{figure}
Here the resonance peaks are absent, because the diagrams
\texttt{Dgg\_sS\_vertFg}, \texttt{Dgg\_sS\_vertSg} and \texttt{Dgg\_sS\_vertSSg}
are parity-conserving. Furthermore, the SUSY corrections to the parity violating
asymmetry are one order of magnitude smaller than the corrections to the
parity-even observables, because it only gets contributions from interferences
of oder $\order(\alpha\alpha_s^2)$.

By integrating the differential cross sections \eqref{eq:res:asymmetries} over
$M_{t\bar t}$ we obtain cross sections for producing $t\bar t$ pairs with
arbitrary invariant mass. We define the cross sections $\sigma_{\text{tot}}$,
$\sigma_{LL}$ and $\sigma_{PV}$ by
\begin{equation}
   \sigma_{\text{tot}/LL/PV}
  =\int_{2m_t}^{M_{pp}}dM_{t\bar t}
   \frac{d\sigma_{\text{tot}/LL/PV}}{dM_{t\bar t}}
  \eqpunct,
\end{equation}
where $m_t$ is the top mass and $M_{pp}=\unit{14}{TeV}$ is the invariant mass of
the proton-proton system. Again, the leading order and SUSY contributions are
indicated by superscripts `LO' and `SUSY', respectively.
\mref{Table}{tab:res:cross_sections} summarises our results for these
cross sections.
\begin{table}
  \centering
  \begin{tabular}{l*{5}{Mr@{.}l}}
    \hlx{hhs[2pt]}
      & \multicolumn{2}{Mc}{\sigma^{\text{LO}}_{\text{tot}}}
      & \multicolumn{2}{Mc}{\sigma^{\text{SUSY}}_{\text{tot}}}
      & \multicolumn{2}{Mc}{\sigma^{\text{LO}}_{LL}}
      & \multicolumn{2}{Mc}{\sigma^{\text{SUSY}}_{LL}}
      & \multicolumn{2}{Mc}{\sigma^{\text{SUSY}}_{PV}} \\
    \hlx{h}
    1a & 331&8 &  +8&4 & 91&99 & +1&75 & +0&063\\
    1b & 288&3 &  +6&9 & 79&50 & +1&72 & +0&020\\
    2  & 266&3 &  -1&0 & 72&84 & -0&34 & +0&125\\
    3  & 289&5 &  +6&9 & 80&08 & +1&73 & +0&043\\
    4  & 307&6 &  +7&1 & 85&62 & +1&64 & -0&018\\
    5  & 332&1 & +19&8 & 93&48 & +5&16 & +0&111\\
    6  & 310&8 &  +7&4 & 85&74 & +1&68 & +0&053\\
    7  & 284&9 &  +3&5 & 78&04 & +0&78 & +0&029\\
    8  & 271&2 &  -1&9 & 74&08 & -0&64 & +0&035\\
    9  & 263&6 &  +6&9 & 72&32 & +1&82 & +0&019\\
    \hlx{hh}
  \end{tabular}
  \caption{\label{tab:res:cross_sections} Numerical results for the integrated
  $t\bar t$ cross section and asymmetries. The numbers in the left column refer
  to the labelling of the Snowmass benchmarks in \cite{Snowmass02}. The
  superscripts `LO' and `SUSY' indicate leading order results and SUSY
  corrections, respectively. The cross sections are given in pico-barns (pb).}
\end{table}
For both, $\sigma_{\text{tot}}$ and $\sigma_{LL}$ we see that the SUSY
corrections typically make up 2\% of the leading order results. However, they
can be as big as 5\% in the case of benchmark 5 and as small as 0.3\% in the
case of benchmark 2.

\begin{figure}
  \centering
  \includegraphics{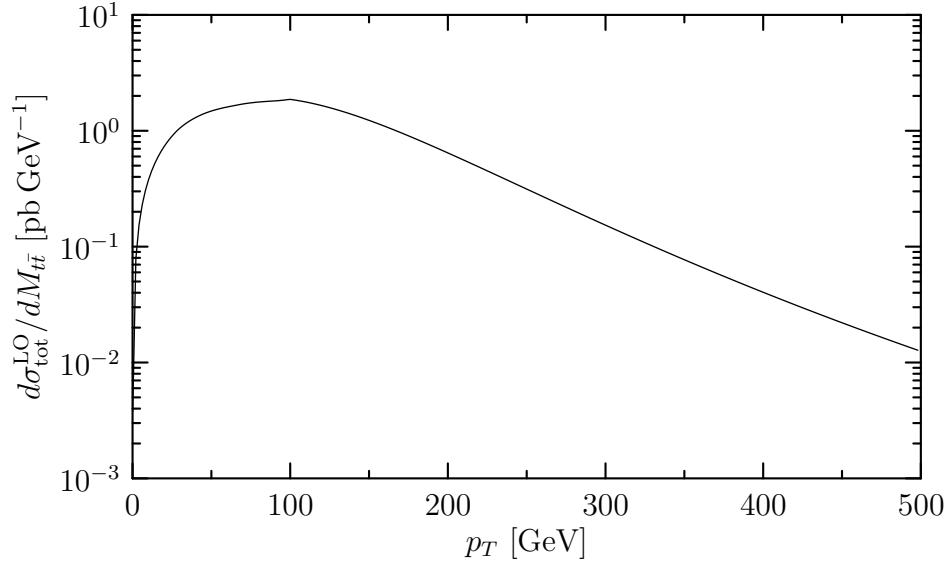}
  \caption{\label{fig:res:tree_tot_pt} Leading order results for the transverse
  momentum differential cross section.}
\end{figure}

\begin{figure}
  \centering
  \includegraphics{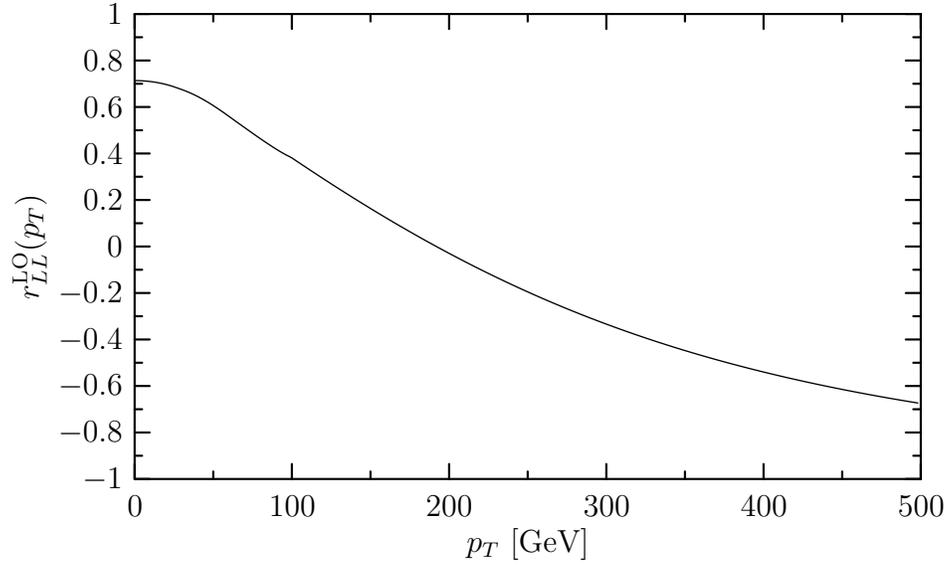}
  \caption{\label{fig:res:tree_al_pt} Leading order results for the transverse
  momentum $LL$ asymmetry.}
\end{figure}

\begin{figure}
  \centering
  \includegraphics{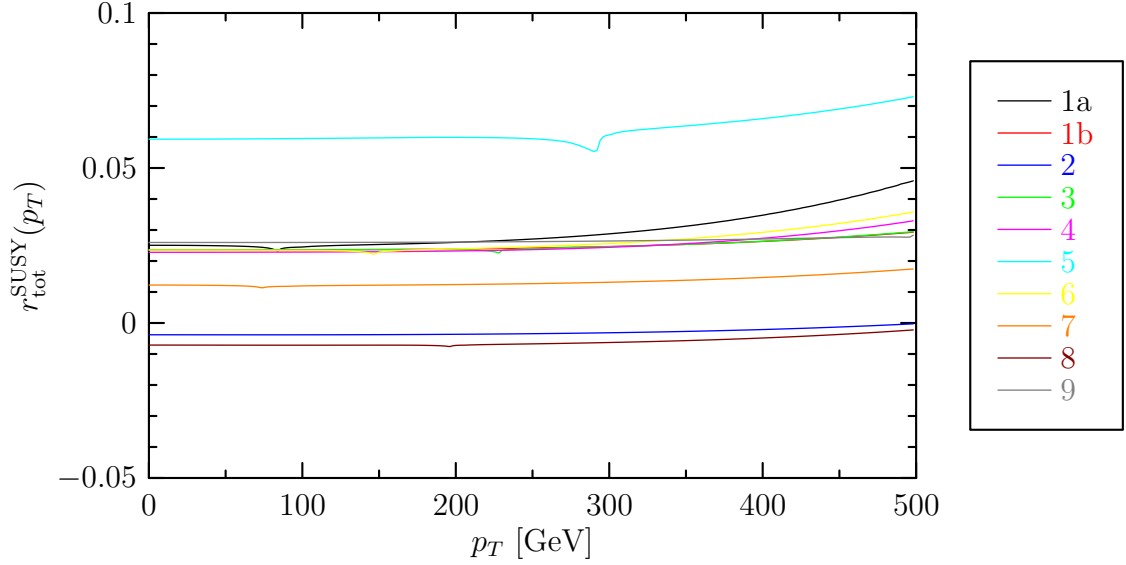}
  \caption{\label{fig:res:tot_pt} SUSY corrections to the transverse momentum
  differential cross section for the Snowmass benchmarks. The numbers in the
  legend refer to the labelling of the benchmarks in \cite{Snowmass02}.}
\end{figure}

\begin{figure}
  \centering
  \includegraphics{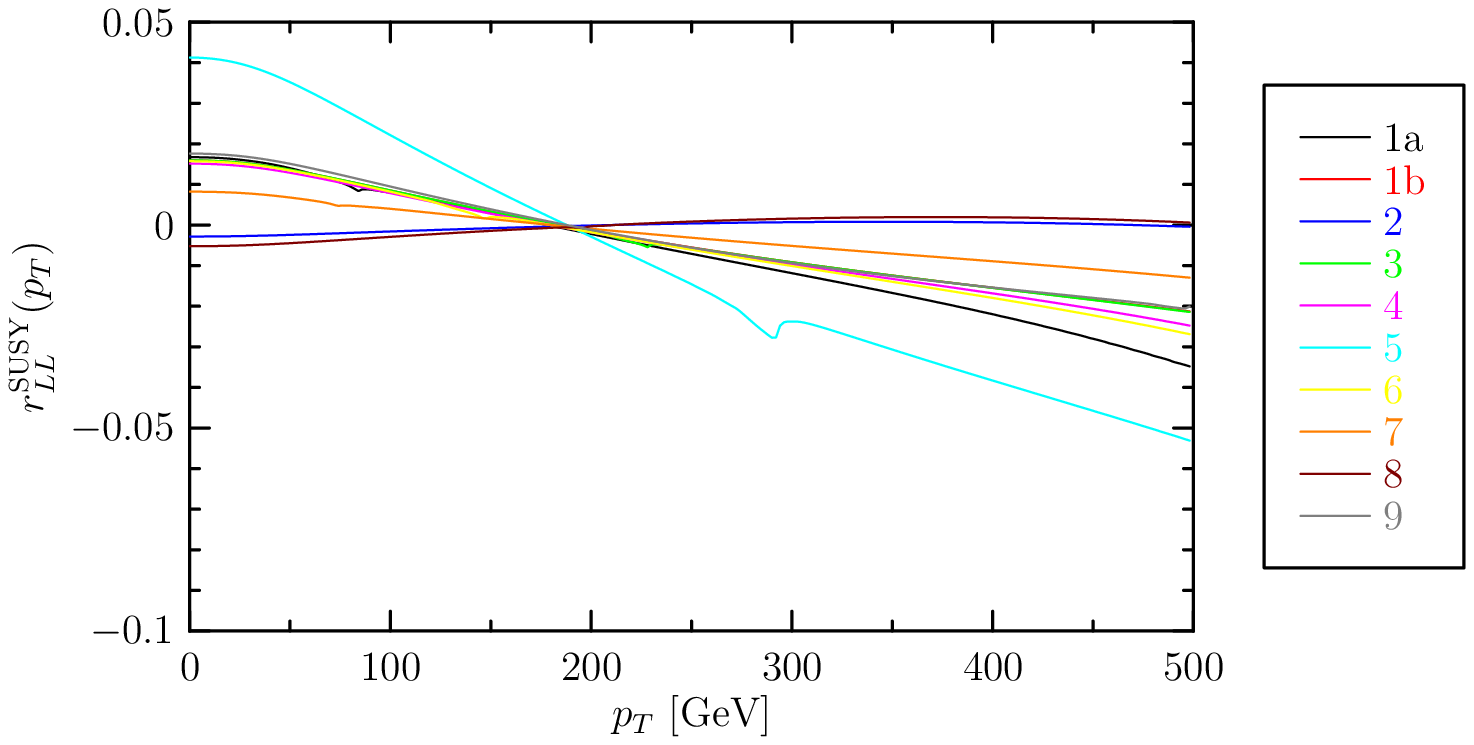}
  \caption{\label{fig:res:al_pt} SUSY corrections to the transverse momentum
  $LL$ asymmetry for the Snowmass benchmarks. The numbers in the legend refer to
  the labelling of the benchmarks in \cite{Snowmass02}.}
\end{figure}

\begin{figure}
  \centering
  \includegraphics{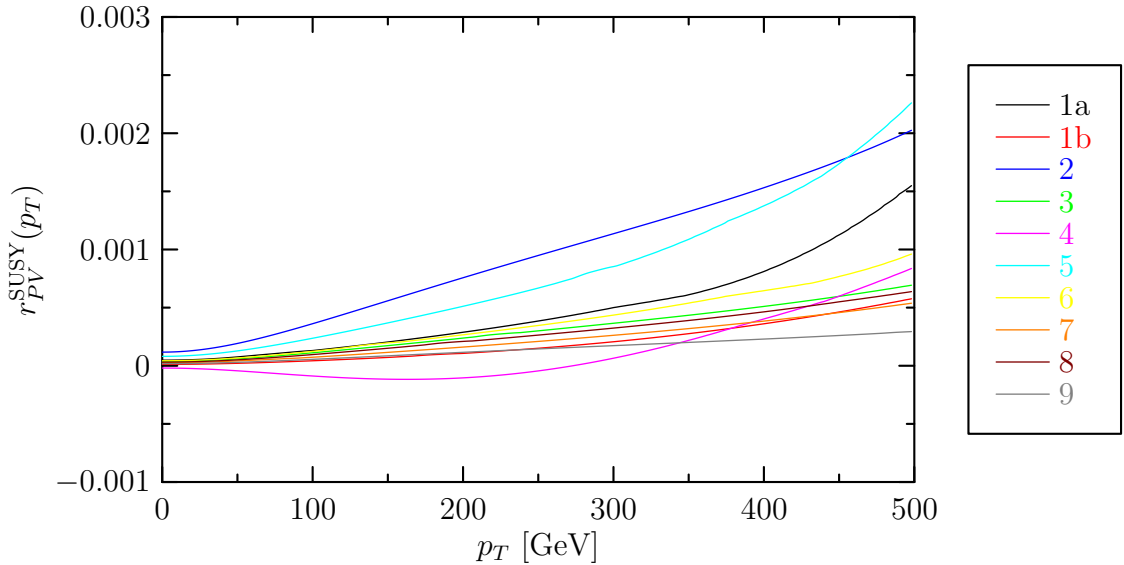}
  \caption{\label{fig:res:pv_pt} SUSY corrections to the transverse momentum
  $PV$ asymmetry for the Snowmass benchmarks. The numbers in the legend refer to
  the labelling of the benchmarks in \cite{Snowmass02}.}
\end{figure}

Experimentally it is often more convenient to parametrise the $t\bar t$
production cross section by the transverse momentum $p_T$ of the top quark.  For
the transverse momentum differential cross section
$d\sigma_{\lambda_1\lambda_2}/dp_T$ we define the total differential cross
section $d\sigma_{\text{tot}}/dp_T$, the asymmetries $d\sigma_{LL}/dp_T$ and
$d\sigma_{PV}/dp_T$ and the ratios $r^{\text{LO}}_{LL}(p_T)$,
$r^{\text{LO}}_{PV}(p_T)$, $r^{\text{SUSY}}_{\text{tot}}(p_T)$,
$r^{\text{SUSY}}_{LL}(p_T)$ and $r^{\text{SUSY}}_{PV}(p_T)$ in analogy to
\eqref{eq:res:asymmetries} and \eqref{eq:res:ratios}. Our results for these
quantities are shown in \mrefs{figures \ref{fig:res:tree_tot_pt} to
\ref{fig:res:pv_pt}}.
We note here that the ``resonance peaks'' and ``troughs'' from the 
thresholds for scalar particle exchange are smoothed out by the
 phase-space integration which means that $M_{t\bar{t}}$ is a far better
 variable in which to analyse the data in order to extract information on
the SUSY parameter set, although we note that  some of the benchmarks
give rise to an enhancement of the differential cross-section of up to 7\%
at large $p_T$.


%% file: summary.tex
\section{Conclusions}
\label{sec:summary}
We have calculated the complete MSSM corrections to the cross-section
for $t-\bar{t}$ production at LHC. The calculation has been set up in terms of
a set of prototype Feynman graphs for the partonic-level helicity 
matrix-elements. In this way the prototypes can be used for any set of
 SUSY parameters by  inserting the corresponding couplings and internal
 masses into the prototypes.

We have obtained numerical results for the ten Snowmass benchmark sets using
the CTEQ PDF's. We find a considerable variation of the effects of the
 one-loop SUSY corrections between the various benchmarks. The benchmark
 giving the largest correction is benchmark 5, which is a super-gravity
 model with small $\tan\beta=5$ and a large negative tri-linear coupling,
$A_0=-1000$ GeV. These large corrections can be understood from the fact
that this large tri-linear coupling generates a light stop mass (258 GeV)
thereby enhancing graphs involving a stop mass inside the loop. This gives 
an enhancement of 6\% in the total production cross-section.

Whereas the corrections for the other benchmarks are somewhat smaller,
they are usually around 3\% and therefroe comaparable to the 
weak corrections calculated by Bernreuther et. al. \cite{bernreuther}
and Kuhn et. al. \cite{kuhn}. Note that whereas the weak corrections
reported in \cite{bernreuther,kuhn} {\it decrease} the prediction
for the cross-sections, the SUSY corrections have a {\it positive} sign for 
most of the benchmarks considered.

Statistically, we expect these events to be easily detectable given the
 anticipated yield of order $10^7$ events over the period of running of LHC.
We have found similar corrections in the asymmetry ratios defined in 
 in Eqs.(\ref{eq:res:asymmetries}) and (\ref{eq:res:ratios}). For these 
asymmetries we also expect cancellation of systematic errors arising from 
uncertainties in incoming parton fluxes, so that these corrections of
order 3\% would exceed the statisical errors by  a factor of $\sim \, 100$.

Given corrections of such significance, it is reasonable to 
assume that corrections in the differential cross-sections will also
be detectable (provided sufficiently large bins are taken).
We have therefore plotted the differential cross-sections with 
respect to the invariant
mass $M_{t\bar{t}}$ of the $t-\bar{t}$ system and also with respect to the
transverse momentum $p_T$ of th $t$-quark. In the former case,
the differential cross-sections display an interesting structure with
 peaks and/or troughs corresponding to thresholds for scalar particle exchanges
in the gluon fusion process.

We have also determined the SUSY contribution to the parity odd helicity
asymmetry. This receives only contributions from the supersymmetric partners
 in the weak-interaction sector, which are suppressed relative to the SQCD
corrections by ${\cal O}(\alpha_W/\alpha_s)$. It would appear, therefore
 that even for benchmark 5, which produces the largest corrections, such
 parity violating  asymmetries will be too small to observe.
\bigskip

{\bf Acknowledgements:}\\
The authors are grateful to Stefano Moretti and Sacha Belyaev for useful
converstaions.